%% file: bare_conf_NDSS2027.tex
\documentclass[conference]{IEEEtran}
\IEEEoverridecommandlockouts
\newcommand{\sniper}{{ICS-Sniper{}}}
\usepackage[available,functional,reproduced]{ndssbadges}
\ifCLASSINFOpdf
\else
\fi
\usepackage[colorinlistoftodos,textsize=scriptsize]{todonotes}
\usepackage{enumitem}
\usepackage{listings}
\usepackage{multirow}
\usepackage{amssymb}
\usepackage{pifont}
\usepackage{algorithm}
\usepackage{algpseudocode}
\usepackage{hyperref}
\usepackage{tcolorbox}
\usepackage{tikz}
\usepackage{amsmath}
\usepackage{tabularx}
\usepackage{amssymb}
\usepackage{enumitem}
\usepackage{caption}
\usepackage{listings}
\usepackage{xcolor}
\usepackage{soul}

\usepackage[table,xcdraw]{xcolor}

\usepackage{booktabs}

\usepackage{subcaption}
\definecolor{mygray}{rgb}{0.5,0.5,0.5}
\lstdefinestyle{ST}{
    language=Lisp,
    basicstyle=\fontsize{8}{9}\selectfont\ttfamily,
    commentstyle=\color{blue},
    morecomment=[l][\color{blue}]{//},
    morekeywords={PROGRAM, VAR, END_VAR, IF, THEN, ELSE, END_IF, OR},
    numbers=left,
    stepnumber=1,
}
\newcommand*\circled[1]{\tikz[baseline=(char.base)]{
\node[shape=circle,draw,inner sep=0.7pt] (char) {#1};}}
\makeatletter
\newcommand{\linebreakand}{%
  \end{@IEEEauthorhalign}
  \hfill\mbox{}\par
  \mbox{}\hfill\begin{@IEEEauthorhalign}
}
\makeatother

\begin{document}
\title{{\sniper}: A Targeted Blackhole Attack on Encrypted ICS Traffic}

% \author{\IEEEauthorblockN{Gargi Mitra\textsuperscript{1}}
% \and
% \IEEEauthorblockN{Chanyuan Liu\textsuperscript{1}}
% \and
% \IEEEauthorblockN{Pritam Dash\textsuperscript{1}}
% \and
% \IEEEauthorblockN{Elaine Yingao Yao\textsuperscript{2}}
% \and
% \IEEEauthorblockN{Alain Zhiyanov\textsuperscript{3}}
% \and
% \IEEEauthorblockN{Aastha Mehta\textsuperscript{3}}
% \linebreakand
% \IEEEauthorblockN{Karthik Pattabiraman\textsuperscript{1}}
% \linebreakand
% \IEEEauthorblockA{\\\textsuperscript{1}Department of Electrical and Computer Engineering, \textsuperscript{2}Department of Computing and Information Science, \\\textsuperscript{3}Department of Computer Science,\\\textsuperscript{13}The University of British Columbia, Vancouver, Canada, \textsuperscript{2}Cornell University, New York, USA\\Email:\{gargi@ece, chanyuanl@ece, pdash@ece\}.ubc.ca, elainey@cs.cornell.edu, \{azhiyano@student, aasthakm@cs\}.ubc.ca,\\ karthikp@ece.ubc.ca}}	
	

\author{\IEEEauthorblockN{Gargi Mitra\IEEEauthorrefmark{1},
Chanyuan Liu\IEEEauthorrefmark{1},
Pritam Dash\IEEEauthorrefmark{1}, 
Elaine Yao\IEEEauthorrefmark{2},
Alain Zhiyanov\IEEEauthorrefmark{3},
Aastha Mehta\IEEEauthorrefmark{3} and
Karthik Pattabiraman\IEEEauthorrefmark{1}}
\IEEEauthorblockA{\IEEEauthorrefmark{1}Department of Electrical and Computer Engineering,
University of British Columbia,
Vancouver, Canada}
\IEEEauthorblockA{\IEEEauthorrefmark{2}Department of Computing and Information Science, Cornell University, New York, USA}
\IEEEauthorblockA{\IEEEauthorrefmark{3}Department of Computer Science,
University of British Columbia,
Vancouver, Canada}
\IEEEauthorblockA{Email:\{gargi@ece, chanyuanl@ece, pdash@ece\}.ubc.ca, elainey@cs.cornell.edu, \{azhiyano@student, aasthakm@cs\}.ubc.ca,\\ karthikp@ece.ubc.ca}
\thanks{\IEEEauthorrefmark{2}Elaine Yao was a Master's student at the Department of Electrical and Computer Engineering, UBC, when she participated in the research presented in this paper.}
%\thanks{The research was funded by the National Cybersecurity Consortium of Canada.}
}


\IEEEoverridecommandlockouts
\makeatletter\def\@IEEEpubidpullup{6.5\baselineskip}\makeatother
\IEEEpubid{\parbox{\columnwidth}{
		Network and Distributed System Security (NDSS) Symposium 2027\\
		22--26 March 2027, Seoul, Republic of Korea\\
		ISBN 978-1-970672-09-1\\  
		https://dx.doi.org/10.14722/ndss.2027.232779\\
		www.ndss-symposium.org
}
\hspace{\columnsep}\makebox[\columnwidth]{}}

\maketitle

\begin{abstract}
Modern industrial control systems (ICS) increasingly host their Supervisory Control and Data Acquisition (SCADA) services in the cloud to reduce the costs of large-scale automation. To protect site-SCADA communications, ICS operators commonly use VPN tunneling and standard security practices. We show that, despite these security measures, an on-path Internet adversary can disrupt ICS operations without infiltrating the ICS perimeter, breaking encryption, or knowing the control logic. 
We present \sniper{}, a targeted blackhole attack that analyzes the VPN traffic metadata (sizes, direction, timing of packets) to identify narrow time windows, called \textit{critical superperiods}, during which the site-SCADA traffic would likely contain highly critical commands or data. Post analysis, in a subsequent operational cycle, \sniper{} drops a small set of payload-carrying packets in the critical superperiods to disrupt the ICS's operations.
We demonstrate three attacks on two realistic modern Secure Water Treatment (SWaT) plant testbeds that can potentially violate the operational safety of the ICS while evading state-of-the-art ICS attack detectors.
\end{abstract}


% \fi
\IEEEpeerreviewmaketitle

%%%%%% Sections start here %%%%%%%%%%%
\input{sections/introduction}

\input{sections/background}

\input{sections/threat-model}

\input{sections/methodology}

\input{sections/experimental-setup}

\input{sections/evaluation}

\input{sections/limitations_countermeasures}

\input{sections/relwork}

\input{sections/conclusion}

\section*{Acknowledgements}
This research was supported in part by the National Cybersecurity Consortium of Canada (NCC),  the Natural Sciences and Engineering Research Council of Canada (NSERC) and the Digital Research Alliance of Canada (DRAC).  
%%%%%% End of sections %%%%%%%%%%%%%%%%

\section{Ethics Consideration}
% We have made our best efforts to conduct our research ethically. We highlight our considerations at specific research steps.
Our ethical considerations at each research step are outlined below.

{\bf Stakeholders.} This research was conducted following a stakeholder-based ethics analysis. We identified several key stakeholders who could be impacted by our research process and its publication:
\begin{itemize}
    \item The Research Team: The authors on this paper.
    \item The ICS: The plant being targeted in the attack.
    (i) Our reconstruction of the superperiodic structure for the SWaT plant at SUTD might raise concerns about threats to the physical plant. However, the original plant's OT system is not connected to the Internet. Therefore, an adversary cannot leverage our work to attack the physical plant.

    (ii) Our attacks shown in \S\ref{sec:evaluation} have been performed only on an emulated testbed that contains no physical components.

    \item Users: The end users served by the plant.
    \item The Security Community: Our peers who build upon and validate security research.
    \item Malicious Actors: Those who might attempt to exploit the vulnerabilities we discovered.
\end{itemize}

{\bf Risk vs. beneficence of our research.} We have considered the Menlo Report principles~\cite{kenneally2012menlo}, particularly “Beneficence”~\cite{kohno2023ethical}, balancing research benefits with privacy protection. While our research introduces some risk that adversaries might launch the attacks exposed in this work, we believe that the benefits of revealing the information for building secure ICS solutions outweigh the potential threat.
% {\bf Vulnerability disclosure.} Following best practices as described in the Menlo Report~\cite{kenneally2012menlo}, we have disclosed the identified vulnerabilities to ??.

{\bf Other considerations.} Our work involves no human subjects or personal data, and all experiments were conducted on isolated local or cloud setups that did not affect other users.

\section{LLM Usage Considerations}
No new text or image in this paper has been generated using LLMs. However, we have used LLM (Claude Sonnet 4.6) to check the grammar and consistency of the text.



%\bibliographystyle{IEEEtran}
\bibliographystyle{IEEEtran}
\bibliography{references}
%
% \end{thebibliography}

% \newpage
% \appendix
% \input{appendix}
% \clearpage
% \input{artifact-appendix}

\appendices                % multiple lettered appendices (IEEEtran)
\section{}                 % empty title: heading reads just "APPENDIX A"
\input{appendix}
\clearpage
\let\appendices\relax      % artifact-appendix.tex calls \appendices again; stop it restarting at A
\input{artifact-appendix}

\end{document}

%% file: sections/introduction.tex
\section{Introduction}\label{sec:introduction}
Industrial control systems (ICS) automate critical processes such as power generation, water treatment, and manufacturing~\cite{otinternet}. Attacks on these systems cause service outages, equipment damage, or safety hazards, resulting in productivity losses in downstream processes, civilian inconvenience, and reputational and financial losses for the parent organization~\cite{canadaattack,texasattack}. At the core of an ICS is the operational technology (OT) network, comprising Programmable Logic Controllers (PLCs), sensors, and actuators. In large plants, the physical process is divided into sub-processes that operate in tight synchronization to ensure safe and reliable operation. Each sub-process comprises multiple PLCs, one of which is designated the gateway PLC and aggregates that sub-process's data. A central Supervisory Control and Data Acquisition (SCADA) software facilitates cross-sub-process synchronization by collecting data from gateway PLCs, serving cross-sub-process data requests, and issuing control commands. Timely delivery of these messages is critical for the physical process.

% At the core of an ICS is the operational technology
% (OT) network, comprising several Programmable Logic Controllers (PLCs), sensors, and actuators. 
% In large plants, the physical process is broken down into smaller sub-processes. The sub-processes must operate in a tightly synchronized manner to ensure safe and reliable operation. A central Supervisory Control and Data Acquisition (SCADA) software facilitates this synchronization by collecting data from PLCs, serving cross-sub-process data requests, and issuing control commands. Since the correct physical behavior of the ICS depends on the timely delivery of these messages, delays or disruptions in communications can push the system into unsafe states, causing service outages, equipment damage, or safety hazards, making these OT networks lucrative targets for attacks for both 
 %\karthik{I think we should reduce the background info and get to the point quickly.}\gargi{Shortened it}
\begin{figure}[t]
    \begin{center}
    \includegraphics[scale=0.45]{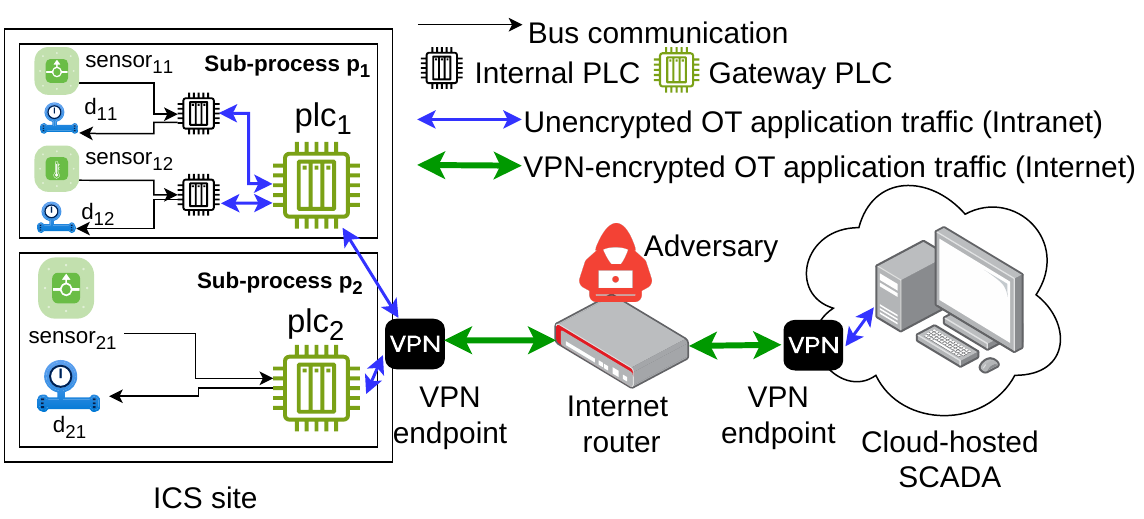}
    % \captionsetup{justification=centering}
    \caption{A simple example of a modern OT network with cloud-hosted SCADA, and the position of \sniper{}.}
    \label{fig:ics-intro}
    \end{center}
\end{figure}

Historically, OT networks were confined to private intranets, minimizing exposure to remote adversaries. However, as industrial plants scaled and operations became geographically distributed, on-premise SCADA systems and Intranet connectivity across the plant grew increasingly inconvenient and expensive~\cite{awscloud}. This has driven a gradual shift toward hosting SCADA systems on cloud platforms, offering cost savings, remote monitoring capabilities to operators, and built-in resilience for power and network connections~\cite{awscloud,vtscada}. Cloud technologies are therefore expected to become critical to OT infrastructure in the coming years. 

% Historically, OT networks were designed to be confined to private intranets, minimizing exposure to remote adversaries. However, as industrial plants scaled up and began spanning larger geographical areas, the increasing demand for distributed computational resources, network connectivity, and management personnel made on-premise SCADA systems increasingly expensive.
% Moreover, strict isolation hindered remote monitoring, maintenance, and scalability, which proved to be inconvenient, especially in sectors like water and energy, where operations are distributed geographically, often in remote areas~\cite{awscloud}.

% These limitations have driven a gradual shift toward hosting SCADA services on cloud platforms,
% which provides cost benefits, enables remote access to process data, supports large-scale analytics and automation, and provides resilience against benign faults through built-in redundancy of hardware, power, and network connections~\cite{awscloud,vtscada}.

% \textbf{Are modern OT networks adequately secure?} 
% Cloud-hosted SCADA also increases the exposure of an ICS to remote adversaries, such as a man-in-the-middle (MITM), which can observe the communication between ICS sites and the SCADA.
Migrating the SCADA to the cloud necessitates site-to-SCADA communications to traverse over the public Internet. To protect these communications from remote adversaries, including Man-In-The-Middle (MITM) attackers, the US NIST and AWS make four recommendations for ensuring the security of Internet-connected OT networks~\cite{nistguidelines, awssecurity}:
(i) access control and message integrity through endpoint authentication and cryptographic mechanisms,
(ii) implementing secure access control with firewalls and intrusion detection systems (IDS),
(iii) ensuring communication confidentiality through Virtual Private Networks (VPNs) tunneling, and
(iv) maintaining high availability by employing early detection of volumetric Denial of Service (DoS) attacks and using redundant OT components.

\textbf{A novel targeted DoS attack.} 
We show that despite these security measures, the migration of the SCADA to the cloud
makes ICSes susceptible to {a new class of Denial of Service (DoS)  attacks} that exploit the expanded attack surface of modern ICSes. %We present a systematic methodology to mount such attacks. 
We introduce \sniper{}, a novel \textit{targeted} blackhole DoS attack that breaks the synchronization among sub-processes by selectively dropping a \emph{small number of critical OT messages} communicated between the cloud-hosted SCADA and on-premise PLCs, thereby causing significant disruption and damage to the ICS. To do so, 
\sniper{} neither infiltrates the ICS network nor breaks traffic encryption to access the payload. Instead, \emph{it relies solely on the metadata of VPN-encrypted plant-to-SCADA traffic.}

\sniper{} operates from an on-path network vantage point outside the plant perimeter, and does not know about the plant's control logic (see Figure~\ref{fig:ics-intro}).  
Yet, it can identify critical packet sequences within a narrow time window, and can cause disruptions by dropping 20\% or 100\% of packets in those windows, depending on the communication protocol used.
% \aastha{Is this 20\% number up to date?}\gargi{Updated}
This enables the adversary (e.g., rogue insiders in an Internet Service Provider (ISP), and enemy nation-state actors with ISP-level access) to perform the attack stealthily, evading conventional DoS attack detectors.
% \aastha{Why would a nation-state actor attack its own infrastructure??}
% \aastha{We might want to name a different type of attacker than enemy nation-state actor. An enemy actor would have to go through many hoops to first infiltrate a country's internet infrastructure to reach the ICS. Can we name some other adversary who might already be inside the network?}\gargi{Done. Please check.}

\sniper{} employs a novel encrypted traffic analysis (ETA) technique~\cite{coull2025traffic} %\karthik{Citation?}\gargi{Added a generic citation to ETA so that reviewers don't think that we just plugged in existing web traffic analysis techniques.} 
to identify the critical packet sequences solely from traffic metadata, such as the sizes, timing, and direction of transmitted packets. 
% This technique is fundamentally different from those used in the context of web privacy and anonymity attacks~\cite{wang2014effective}, 
% Unlike the web domain, the \sniper{} adversary has no knowledge of the internals of the ICS. 
For this purpose, it exploits two structural properties of ICS communication. First, each sub-process follows a state-transition model, where a state represents a unique combination of the mechanical device configurations. A sub-process transitions between states in response to messages from remote sub-processes. Second, each PLC in a given state exchanges a characteristic, periodically repeating message sequence with the SCADA, producing periodic, bursty traffic, with burst patterns changing detectably when a state transition occurs. 
Therefore, dropping packets close to the state transition time of a PLC breaks synchronization among sub-processes, pushing the ICS into an unsafe state. 

\sniper{}'s core task is therefore to predict when these state transitions (or burst pattern changes) are imminent. 
The key challenge in inferring this is that often multiple gateway PLCs share a common VPN gateway, producing interleaved traffic streams that obscure these per-PLC periodic burst patterns. \sniper{} addresses this by decomposing the observed VPN traffic into ``superperiods''---repeating composite patterns that capture fixed multiples of each constituent PLC's periodic traffic patterns, from which it predicts upcoming state transition timings. \sniper{} identifies the superperiod by using phase folding~\cite{zucker2015detection}, a technique from astronomical signal processing, which is particularly efficient in signal period determination in the presence of noise and frequently changing burst patterns. %\gargi{Added a point on the phase-folding technique.} 
For adapting this technique to ICS traffic, \sniper{} incorporates protocol-level knowledge of ICS standards (e.g., Modbus, ENIP), such as packet ordering, packet size ranges, and message transmission sequence, into its ETA technique.
%\karthik{What's the protocol-level knowledge it exploits for constructing superperiods?}
During the attack, ICS-Sniper drops payload-carrying packets in the superperiod immediately preceding a predicted transition and during the transition itself, maximally disrupting the physical process at the most opportune time.

To evaluate \sniper{}, we built two complementary validated testbeds based on the six-stage Secure Water Treatment (SWaT) plant~\cite{swat-testbed}. The first testbed, which we refer to as ``S-SWaT'', contains software-simulated PLCs, enabling fast prototyping. The second testbed, referred to as ``H-SWaT'', uses real industry-standard PLCs for reproducing the timing and communication characteristics of real deployments.
% \aastha{The names $Testbed_1$ and $Testbed_2$ are uninformative. Maybe replace with something like ``S-SWaT'' and ``H-SWaT'' for software and hardware testbed.}\gargi{Done}
For both testbeds, we hosted the SCADA on a cloud VM, which communicates with PLCs over an OpenVPN tunnel~\cite{openvpn}.
% While we used the same control and communication logic as that of the original SWaT testbed, we migrated the SCADA to the Amazon EC2 instances, which is consistent with other cloud-hosted SCADA deployments on AWS~\cite{awscloud}.
% \aastha{Didn't you move from AWS to an internal cloud infrastructure?} \gargi{Not yet. We decided to migrate it after paper submission.}
% The PLCs reside in an organizational network and communicate with the SCADA over an OpenVPN~\cite{openvpn} tunnel. 
% The traffic is routed through another cloud VM that simulates a compromised router and runs \sniper{}. 
The testbeds support industry-standard communication protocols, configurable communication periods and network conditions,
% Both testbeds incorporate
network delay-tolerance mechanisms, and an ensemble~\cite{wolsing2024ensemble} of DoS attack detectors. To our knowledge, these are the first testbeds providing
% where on-premise PLCs communicate with
a cloud-hosted SCADA over VPN.

{%\color{red}
\textbf{Key results.} %\karthik{Can we also describe the ICS configuration in 1 sentence first, and how we ran the experiment.}\gargi{Added}
1) {\sniper} could infer the superperiod in the plant with an accuracy of $100\%$ in both testbeds.
% \aastha{Evaluation uses the word accuracy. Why is it precision here?}
% \aastha{you mean both testbeds?}
% 2) Compared to the entire operational cycle, each superperiod occupies only a small time window. \sniper{} targets only $20\%$ of the packets during two such superperiods, which is stealthier compared to traditional volumetric DoS attacks that gradually flood the entire bandwidth or drop packets en masse~\cite{gasser2017amplification,airehrour2016securing,deepstrike} for minutes to hours.
% \aastha{Need citations for various types of DoS detectors here.} 
% Therefore, it is difficult for volumetric DoS detectors to detect \sniper{} in real time, and leaves little time for other attack detectors (for e.g. rule-based detectors) to detect the attack before any major damage occurs. 

% \aastha{Is this a top-level result or is this a low-level result in the context of the paper's key contributions?}
2) \sniper{} dropped critical packets at predicted superperiods for 14\,minutes in S-SWaT. One attack delayed the start of the distribution of purified water by 15\,minutes, while another caused a loss of purified water for 9\,minutes 11\,seconds. Similarly, a one-minute packet drop in H-SWaT caused a two-minute delay in the start of purified water distribution.
% \aastha{Why is the second attack on H-SWaT not mentioned}
% \aastha{The enumeration is inconsistent. Also, what are Process delay attack and UF-depletion attack? They appear out of the blue.}\gargi{Updated the results}
% \aastha{The description of the attacks assumes knowledge of the testbed setup, e.g., ``clean-water pumping'', ``feed water'', ``backwash valve'', which we do not present in the intro. Remove the details and present the attack impact in abstract.}
Such disruptions would adversely affect downstream plants relying on the supplied water or inconvenience hundreds of users. 

%\karthik{If we can, we should be more quantitative about this. How many users? What's the period of disruption?}\gargi{Period of disruption: 14 mins, mentioned in \textbf{(2)}}
%\karthik{It would be good to put a cost on this, if possible.}
3) No \sniper{} attack drew alerts from both detector
types {before it impacted the ICS performance}. 
%\karthik{Mention the attack detection techniques and what they do, again in 1-2 sentences. Also, what does timely manner mean?}\gargi{Attack detection techniques are described in the paragraph before Key Results}
These results collectively show that {\sniper} can achieve high-impact, low-visibility attacks even in secure ICS deployments.
}

\textbf{Contributions.}
% In summary, as ICSes move towards could-connected OT networks, we uncover an emerging threat.
\emph{To the best of our knowledge, we are the first to uncover an emerging threat that could affect cloud-connected OT networks and cause significant damage to an ICS. }%\karthik{Can we make this stronger? Perhaps say something like "To the best of our knowledge, we are the first to ..."}
Concretely, we make four contributions in this work.
% {\color{red}\textbf{Need to add section references}}
\begin{enumerate}[wide, labelwidth=!, labelindent=13pt]
    \item Present \sniper{}, a \textit{novel and practical ETA-based targeted DoS attack} that requires no knowledge about system internals, cryptographic secrets, or traffic payloads (\S\ref{sec:methodology}).
    \item Provide two \textit{realistic testbeds} simulating a modern water treatment plant for evaluating our attack as well as for conducting future security experiments (\S\ref{sec:experimental-setup})\footnote{The artifacts for this paper are available on \href{https://doi.org/10.5281/zenodo.23737011}{Zenodo}. These include detailed instructions for reproducing S-SWaT and H-SWaT, network traces captured from our testbeds, and \sniper's code.}.
    % \aastha{Need to connect the implementation of the testbed connect to the contribution of the new ETA-based attack. Otherwise, the testbed stands out of place.}
    \item Present a \textit{comprehensive evaluation} of \sniper{} to highlight its efficacy in disrupting plant operations, and generalizability across various PLC configurations, communication protocols, and network conditions~(\S\ref{sec:evaluation}).
    \item Highlight issues with existing DoS attack detectors, discuss countermeasures and challenges in adapting them to secure OT networks against {\sniper} %\karthik{Perhaps also say we highlight a common issue with all existing DoS based detectors.} 
    (\S\ref{sec:countermeasures}).
\end{enumerate}

% Our work highlights the need for stronger protections in modern ICSes beyond the NIST recommendations.
% We discuss countermeasures, e.g., network-layer traffic shaping, that have been studied in other security contexts~\cite{sabzi2024netshaper} and the challenges involved in adapting them to secure OT networks against {\sniper} (\S\ref{sec:countermeasures}), and provide actionable insights for securing next-generation industrial control systems. 

%% file: sections/background.tex
\section{Background}\label{sec:background}
% \aastha{The purpose of this background is unclear. Connect a bit to the goal of the paper or what is coming up in the next sections.}\gargi{Please check now}
% We briefly discuss the growing adoption of cloud-hosted SCADA in modern ICSes and their evolving security landscape to motivate the need for analyzing their vulnerabilities. 

\subsection{Cloud-hosted SCADA}
The Supervisory Control and Data Acquisition (SCADA) system helps in centralized monitoring and control of geographically dispersed industrial assets~\cite{nistguidelines}. PLCs at a site report process telemetry upstream to the SCADA, enabling plant operators to remotely monitor operations and detect anomalies. The SCADA transmits coordination data downstream to synchronize activities across sub-processes. 

% The SCADA acquires data from the sub-processes in real time,   
% and distributes them in real time to remotely located sub-processes, thereby facilitating synchronization among remotely located devices. It also presents the data to the plant operators for monitoring the plant's health and detecting anomalies.

% Cloud-hosted SCADA is the latest stage in the evolution of ICS OT networks. 
The migration of SCADA from the plant site to the cloud is driven by cost and operational benefits.
% Migrating the SCADA from the plant site to the cloud offers several benefits such as lower infrastructure and maintenance costs, improved reliability, remote visibility into plant data, and on-demand resource provisioning that facilitate flexible payment models~\cite{awscloud}.
This is especially valuable for industries managing large-scale infrastructure at remote sites with limited staff. Cloud providers further reduce operational downtime through high-availability guarantees. Industrial surveys confirm a recent surge in cloud-hosted SCADA adoption~\cite{sans2024}, with water and wastewater utilities across the USA, Canada, Mexico, Australia, and New Zealand among early adopters~\cite{USA, vtscada, yokogawa}.
% This is particularly helpful for industries that require managing a huge infrastructure at remote sites with limited IT staff. Cloud providers also offer high availability guarantees, which reduce operational downtime in the event of faults.
% Industrial surveys show a surge in the adoption of cloud-hosted SCADA in recent times~\cite{sans2024}.
% Examples of industries that have migrated their SCADA to the cloud include water and wastewater industries in the USA, Canada, Mexico, Australia, and New Zealand~\cite{USA, vtscada, yokogawa}.

% \gargi{Newly added to address reviewers' concerns about the number of PLCs communicating with the SCADA}
\textbf{PLC-SCADA Communications.} Large-scale ICSes employ a hierarchical communication architecture to minimize Internet-bound data flow and limit adversarial exposure to sensitive operational data. %Even in deployments with hundreds of PLCs, 
Direct communication with the cloud-hosted SCADA is not permitted for every device. Instead, if a sub-process has multiple PLCs, it designates a single gateway, either a PLC or a Remote Terminal Unit (RTU), as its sole aggregation point for upstream communication\footnote{RTUs and PLCs use similar communication protocols.} (see Figure~\ref{fig:ics-intro}). The remaining PLCs within that sub-process execute local control logic and communicate exclusively with peer devices and the gateway device over the internal network. The typical number of gateway PLCs within the premises of large plants is $\approx$10–20, and they are typically split across multiple VPN tunnels. For \sniper{}, disrupting one shared tunnel suffices.
% This design confines Internet-facing traffic to one or a few endpoints per sub-process, substantially reducing the attack surface without compromising operational efficiency. 
\textit{%In the \hl{rest} of this paper, 
We focus only on the communication between the gateway PLCs and the SCADA.}
%\karthik{The gateway can also be an RTU, can't it?}
%\gargi{It can. RTUs also use similar protocols, so results won't be different. I will mention it in the footnote.} 

\subsection{Security of Modern OT Communications}
OT communications over the Internet must contend with two classes of issues: (i) delays and drops due to network congestion or variable link speeds, and (ii) security threats such as MITM and DoS attacks. ICS designers address the former through two delay-tolerance mechanisms.
First, legacy protocols such as Modbus~\cite{modbus} and CIP/ENIP~\cite{cip-enip} interface with the TCP/IP stack, relying on TCP retransmission to recover lost or delayed packets~\cite{collantes2015protocols, modbustcp, cip-enip-packet}. These protocols predominantly use synchronous communication, i.e., each PLC issues a new request only after receiving a response or timing out~\cite{modbustcp-packet}.
Second, if messages are lost or delayed, sub-processes continue local computations using the last received data until retransmission succeeds~\cite{krotofil2014cps}.

Legacy OT protocols have been upgraded to secure variants such as Secure Modbus~\cite{secure-modbus-1} and CIP/ENIP-over-TLS~\cite{enip-tls} to protect ICS traffic. VPNs provide an additional layer of confidentiality by obfuscating endpoint addresses as well. ICSes commonly use them to secure site-to-SCADA communication by routing OT traffic between statically addressed VPN gateways~\cite{otvpn-1}. A site-to-SCADA VPN tunnel can be shared by multiple gateway PLCs. These OT VPN tunnels are not shared with the IT network due to security and performance concerns (as per NIST SP 800-82 recommendations)~\cite{nistguidelines}. ICSes are further equipped with anomaly detectors based on process logs~\cite{cheng2019invariant, aoudi2018pasad, wolsing2022simple} and network traffic~\cite{mirsky2018kitsune, lin2018iat}. % to detect attacks and faults.
%\karthik{Do we need to say more about these detectors? Or does it come later?}\gargi{Detectors are detailed in experimental setup}
% Consequently, the VPN does not provide real anonymity from on-path network eavesdroppers similar to that available in personal VPNs~\cite{otvpn-2}.
% the VPNs do not provide anonymity from on-path network eavesdroppers similar to that available in personal VPNs~\cite{}. The OT traffic in such cases is routed through a VPN tunnel connecting two VPN gateways with static IP addresses on either endpoints.
% In general, TLS and VPNs (to the best of our knowledge) do not obfuscate traffic metadata end-to-end, which makes {\sniper} viable even in the presence of these security measures.

\subsection{Phase Folding Technique}\label{sec:phasefolding}
% \sniper{} adapts the phase folding~\cite{zucker2015detection} technique from the astronomical signal processing domain to encrypted network traffic to discern the periodic structure in encrypted OT traffic.
Phase folding is a technique for uncovering hidden periodic patterns in noisy time series data of events. In the context of time series analysis and phase folding, an event is a discrete, instantaneous occurrence of a feature recorded at a specific point in time. Given a time series of events, the technique works as follows. First, it computes the time difference between every possible pair of events, and then tests each difference as a candidate period. For each candidate period T, this operation ``folds'' the timeline onto a single interval of length T, causing events separated by integer multiples of T to overlap along the folds. If T matches the true underlying period, 3 or more recurring events would align consistently after folding. In contrast, incorrect candidate periods would produce dispersed or inconsistent event alignments. 
The number of aligning events produces the alignment score. The candidate yielding the highest alignment score is identified as the true period. In the event of a tie, the one spanning the maximum length of the time series is selected as the true period.

%% file: sections/threat-model.tex
\section{Preliminaries}\label{sec:prelims}
We start with an example of an ICS and then describe how we represent an ICS as a state-transition system, which is crucial for understanding the design of \sniper{} (\S\ref{sec:hlov}).

% \aastha{My sense is that having an example will be useful. However, you might want to use it as a running example to explain the general concepts of local and global state transition models and the scan-cycle jitter, etc. So I would start with the example, then show its state-transition diagram and use it to explain local/global state transition models and the periodicity characteristics. On the other hand, this example seems a bit outdated because it does not explain what is sent to/received from the cloud-based SCADA. Might want to add it to the example.}
%\karthik{I agree with Aastha. Also, do we really need to show it as code or can we show it as a diagram of some sort?}
%\gargi{Moved code to appendix. Only keeping figures for explanation.}

\textbf{ICS example.} 
Consider the 2-stage water treatment plant in Figure~\ref{fig:st-example}, with two sub-processes:  $p_1$ for raw water collection and $p_2$ for chemical treatment, synchronized through a cloud-hosted SCADA. 
Sub-process $p_1$ contains a pump $pmp_1$, a water inlet valve $vlv_1$, a water level sensor $lvl_1$, and a gateway PLC $plc_1$. Sub-process $p_2$ also contains a pump $pmp_2$, a water inlet valve $vlv_2$, a water level sensor $lvl_2$, a flowmeter $flw_2$, and a gateway PLC $plc_2$. The pumps can be either {\tt On} or {\tt Off}, the valves can be {\tt Open} or {\tt Close}, and the water level sensors output a {\tt Low} or {\tt High} value, based on whether the water level is below or above a minimum and maximum threshold, respectively; $flw_2$ outputs the volume of treated water pumped out of $pmp_2$ since the start of the day. Each sub-process also contains a tank ($Tank1$ and $Tank2$) for holding water, and a pipeline connects them both. The plant works for 10 hours a day and must treat W gallons of water within this time, requiring timely coordination between $plc_1$ and $plc_2$ via the SCADA. The communication periods of $plc_1$ and $plc_2$ are 50ms and 40ms, respectively.

% two sensors ($sensor_{11}$ and $sensor_{12}$), and a PLC ($plc_1$). Sub-process $p_2$ also contains two sensors ($sensor_{21}$ and $sensor_{22}$), two devices ($d_{21}$ and $d_{22}$) and a PLC ($plc_2$).

% Each device can be either {\tt ON} or {\tt OFF}.
% Each sensor measures local physical parameters and outputs a {\tt HIGH} or {\tt LOW} value. 
% Each PLC periodically reads local sensor data and relevant remote sub-process data from the SCADA, executes the corresponding control logic, and transmits the updated process parameters to the SCADA, with periodicities $T_1$ and $T_2$ for $p_1$ and $p_2$ respectively.

Figure~\ref{fig:samplecontrolalg} in Appendix~\ref{subsec:control-algo} shows the control logic for $p_1$ ({\tt P1\_Control}) and $p_2$ ({\tt P2\_Control}).
% \aastha{Missing figure ref.}
At the start of the operational cycle, all pumps are off, valves are closed, water level sensors are {\tt Low}, and the flowmeter is set to 0. 
At a scheduled daily time, the PLCs are automatically turned on, and the SCADA starts the operations by setting the global $plantOn$ signal to 1 and sending it to the PLCs, which then open their water inlet valves. Once the water level in $Tank1$ exceeds the minimum level, it turns on $pmp_1$ and starts filling $Tank2$. Once $Tank2$ is full ($lvl_2$={\tt High}), $plc_2$ closes $vlv_2$. At that point, $plc_1$ must also turn $pmp_1$ off within a bounded delay to prevent damage to the pipeline due to additional pressure. Thus, device configurations in $p_1$ are influenced by remote events in $p_2$. Both PLCs read the value of $plantOn$ from the SCADA; PLC $plc_1$ also reads the value of $vlv_2$ from the SCADA over the Internet. The PLCs read the rest of the local sensor values over their local network. To enable this, $plc_1$ periodically sends data requests to SCADA.
% , while both PLCs periodically send the updated process parameters to the SCADA after executing their control logic. 

% devices are {\tt OFF} (lines $4$-$5$ of both algorithms) and both the sensors in $p_1$ output a {\tt LOW} value. When $sensor_{11}$ becomes {\tt HIGH}, $d_{11}$ turns {\tt ON}. When $sensor_{12}$ becomes {\tt HIGH} while $d_{11}$ is {\tt ON}, $d_{12}$ also turns {\tt ON} (lines $12$-$17$ of {\tt P1\_Control}). 

% In $p_2$, device configurations are influenced by remote events in $p_1$ - $d_{21}$ is turned {\tt ON} when $d_{11}$ turns {\tt ON} (lines $9$-$10$ of {\tt P2\_Control}), and $d_{22}$ turns {\tt ON} if $d_{12}$ (in $p_1$) turns {\tt ON} while $d_{21}$ is also {\tt ON} (lines $11$-$13$ of {\tt P2\_Control}). 

% Here, $plc_1$ reads values of $sensor_{11}$ and $sensor_{12}$ over its local network. Likewise, $plc_2$ reads the value of $sensor_{22}$ locally, but obtains the value of $d_{11}$ from the SCADA over the Internet. To enable this, $plc_2$ periodically sends data requests to SCADA, while both PLCs periodically send the updated process parameters to the SCADA after executing their control logic. 

\subsection{ICS as a State Transition System}\label{subsec:ics-state-trans}
%\karthik{Which of these observations are novel or made by us? Seems like most of this section can go to the background if there're none. Otherwise, let's retain and highlight only our observations.}

\begin{figure}[t]
    \begin{center}
    \includegraphics[scale=0.4]{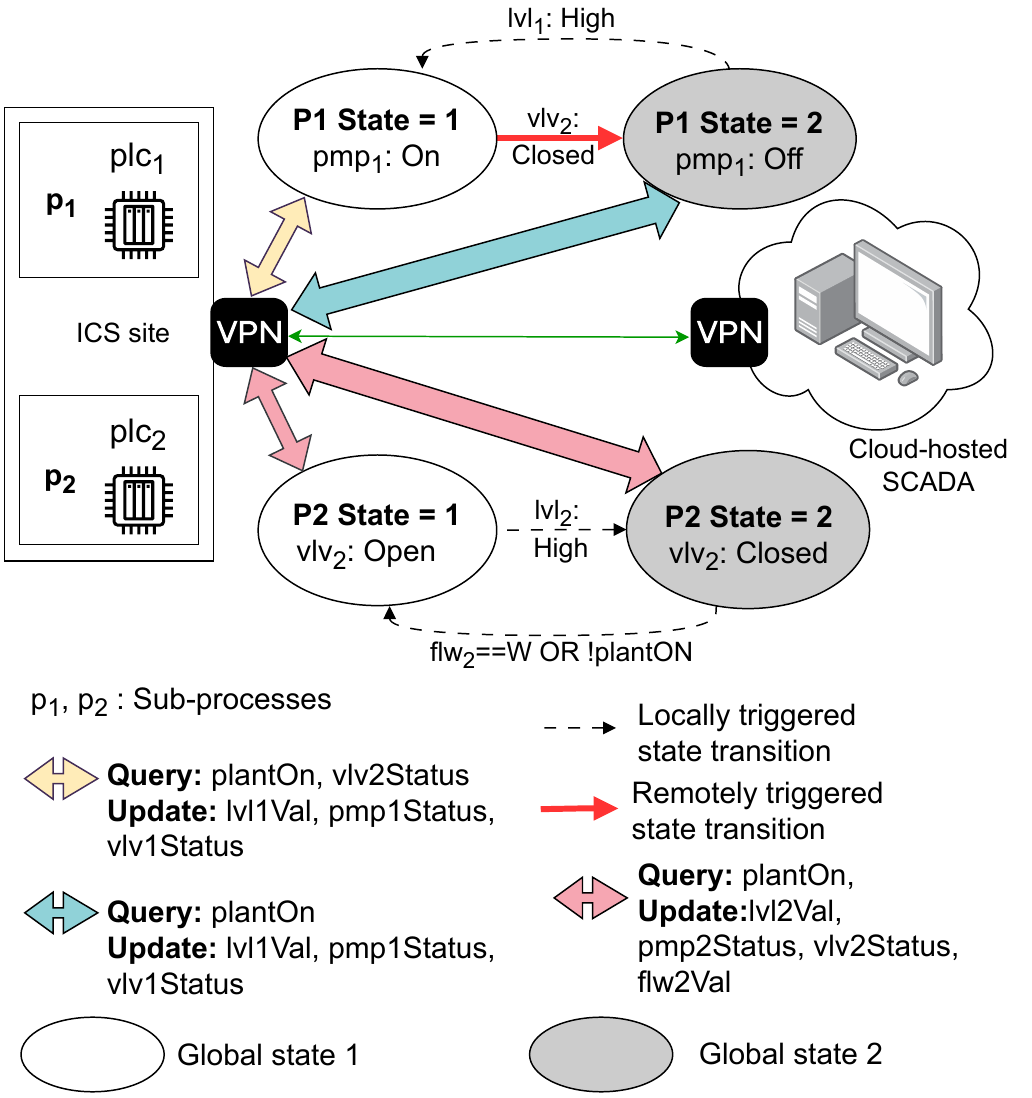}
    % \captionsetup{justification=centering}
    \caption{Operations and communication of the sample ICS (§III) as a state-transition system. 
}
    \label{fig:st-example}
    \end{center}
\end{figure}

\textbf{Local and global state transition systems.} Over a complete operational cycle, each sub-process moves through a sequence of states, each state being a distinct device configuration with its corresponding control logic. A transition occurs whenever the control logic modifies one or more configurations. Because the ICS comprises multiple interacting sub-processes, it also evolves through a series of global states, each a unique combination of sub-process states.
% To summarize, while each sub-process follows its own local state-transition model, the overall ICS process follows a global state-transition model, and any local state transition manifests as a transition in the global state as well. 
Figure~\ref{fig:st-example} shows the state transition systems of our two example sub-processes, with states and transition conditions matching the physical and control conditions described in the example. Note that $p_1$'s transition from State 1 to State 2 is triggered by a remote event, making it a transition of interest to \sniper{}. Crucially, the correctness of global state transitions depends not only on individual sub-processes, but also on the timely exchange of time-critical information between sub-processes.

% Of all the state transitions in this example, the transition of $p_1$ from State 1 to State 2 is triggered by a remote event, and therefore, important to \sniper{}.
% The correctness of the global state transitions depends not only on the independent execution of each sub-process, but also on the timeliness with which sub-processes exchange time-critical information.

\textbf{Safe and unsafe global states.} Safe operation requires all sub-processes to stay synchronized, and only some combinations of sub-process states are safe, as dictated by the system's design, control logic, and timing constraints. In our example, the combinations \{$p_1$ state = 1, $p_2$ state = 1\} and \{$p_1$ state = 2, $p_2$ state = 2\} form the only two safe global states, as indicated in Figure~\ref{fig:st-example}.
Any other combination, for instance, \{$p_1$ state = 1, $p_2$ state = 2\} would be unsafe for the plant, because if the valve in $p_2$ closes but the pump in $p_1$ fails to stop within the required time window, excessive pressure may build up in the pipeline, causing it to burst. Changes in $p_2$ must therefore be relayed to $p_1$ within a bounded time window. 
% The timing dependencies among sub-processes determine how often PLCs should exchange data with the SCADA, thereby determining the overall communication pattern of the system. 
 
\textbf{Periodicity of communications.}
% \aastha{Please compress this section. It is repeating a lot of things already mentioned in the intro.}\gargi{Done}
Each PLC runs two concurrent cycles: a scan cycle that reads inputs, executes control logic, and updates outputs; and a communication cycle that handles all I/O, including \textit{site-to-SCADA communications}. The scan-cycle period is set automatically from the minimum control-logic execution time, whereas the communication-cycle period is configured explicitly by programmers. Since PLC communications directly affect cross-PLC synchronization and demand deterministic timing, they are assigned a fixed period throughout the operational cycle. This is essential for synchronized and predictable system behavior~\cite{allen-bradley-manual}. To tolerate network variability, this period is set slightly above the worst-case communication delay observed during the development and testing stage. For instance, if a PLC sends $x$ messages per cycle, each with an average RTT of $y$ ms, its period is configured as (xy + $\delta_t$) ms, where $\delta_t$ is a jitter margin derived from observed network jitter. If a communication cycle completes early, the PLC remains idle until the period expires before initiating the next cycle.
A PLC's communication period determines its tolerance for communication delays, and hence, the delay tolerance of its state transitions. A delay in scheduled state transition might break inter-sub-process coordination, jeopardizing ICS performance and safety~\cite{nistguidelines}.
%\karthik{This is an important claim and we need a reference to back it up. Or we should refer to our own results.}\gargi{Added citation to NIST guidelines document which mentions this.}

% The time taken to complete these steps is called the scan cycle time. 
% Since the number of inputs, outputs, and control logic varies across states, the scan cycle time also fluctuates. Such variations, known as scan-time jitter, can disrupt synchronization among PLCs. To prevent this, PLC programmers configure a minimum scan time for each PLC, which must be slightly greater than the maximum anticipated scan cycle time of a PLC over the entire operational cycle.
The volume and content of messages sent by a PLC to the SCADA may vary across process states.
% Each PLC in the OT network has its own communication period determined by the time sensitivity of its operations. Over a complete operational cycle, the periodicity of a PLC remains constant, although the messages exchanged with the SCADA might vary across states. Each state contains one or more scan cycles, depending on how long it takes for the control actions in that state to alter device configurations.
% \aastha{This is at odds with our observations right, where the periods are not strictly constant.}
% \aastha{Is this a universal mechanism in real-world ICSes, or is this what we saw in the SuTD testbed and implemented in our own testbed? In either case, would be good to provide a citation.}
% The periodic behavior of PLCs is reflected in their network communications as well. During the input phase, PLCs might need to retrieve data from the SCADA, and after executing the control logic, they transmit updated parameters to the SCADA. The SCADA may also request or push data updates. 
Consequently, each sub-process state, and therefore, each global state, corresponds to a distinct set of network communications. 
In the example shown in Figure~\ref{fig:st-example}, $p_2$ exchanges the same set of messages across both its states, but there is one extra message in the first state of $p_1$ as compared to the second.

\subsection{Characteristics of OT VPN traffic}\label{sec:vpnchar}To characterize PLC traffic over a VPN tunnel, we implemented the example control logic on a real PLC and our cloud-hosted SCADA, and observed the network traffic between them. 
% The network topology of this setup is similar to that of our testbeds, as described in \S\ref{subsec:emulated-swat}. 
Three key observations guide the design of \sniper{}.

\textbf{O1. Number and ordering of packets:} PLC-SCADA communications use either a connected or unconnected protocol variant. In the connected variant, a persistent TCP connection carries each update as a single PLC request and a single SCADA response. In the unconnected variant, each update opens a new TCP connection and comprises a fixed sequence of 10 packets sent in strict order. Under heavy bursts, the gateway's TCP layer may coalesce encrypted packets into one TCP segment, though an adversary can still infer their number and sizes from VPN-layer headers~\cite{openvpn}.
% \aastha{Provide citation of the VPN protocol you used that shows this behavior.}
Across both variants, the request sequence within a state may vary across cycles, but communication is serialized. After each request, the PLC waits for the SCADA's response before sending the next.

\begin{figure}[t]
    \begin{center}
    \includegraphics[width=\columnwidth]{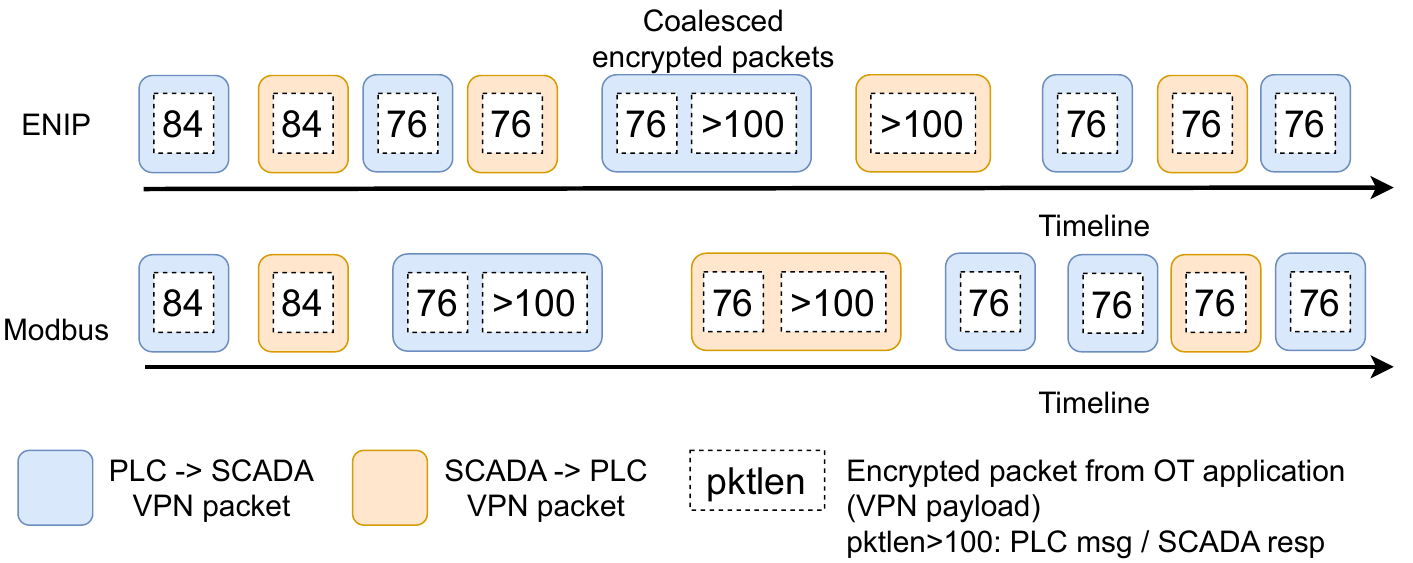}
    % \captionsetup{justification=centering}
    \caption{Traffic patterns of unconnected protocol variants.
    }
    \label{fig:protocol-pkt-seq}
    \end{center}
\end{figure}

\textbf{O2. Packet size sequence:} The sequence of sizes of the 10 encrypted packets in the unconnected variants, mentioned in \textbf{O1}, exhibits protocol-specific patterns, as shown in Figure~\ref{fig:protocol-pkt-seq}. The size of the first packet is always 84 bytes. Of the 10 packets exchanged, 6 are sent by the PLC to the SCADA and 4 in the reverse direction. Only two packets carry the application payload\footnote{The remaining packets are used for connection establishment and teardown.}. For ENIP, these packets are larger than 100 bytes, and for Modbus, they are larger than 85 bytes. The payload packets in the connected variants have similar sizes.

\textbf{O3. Packet size variability:}
Modbus allows aggregating multiple PLC requests into one packet to cut round-trip overhead and latency, which makes packet sizes vary with request grouping. ENIP does not support aggregation but uses symbolic (tag-name) addressing. Since tag names are variable-length strings, packet size depends on the referenced tag name, introducing size variability even without request aggregation. %\karthik{I don't understand this sentence.}\gargi{Rephrased}
% \aastha{Mixing "segments" and "packets" arbitrarily!}
% : \\
% 84, -84, 76, -76, 76, x, -y, 76, -76, 76 (for ENIP)\\
% 84, -84, 76, x, -76, -y, 76, 76, -76, 76 (for Modbus).\\
% Positive segment sizes denote TCP segments sent by the PLC to the SCADA, and negative sizes denote segments sent in the reverse order. Here, x and y represent the packet sizes carrying the application payload, while the remaining packets are used for connection setup and teardown. For ENIP, x, y $>$ 100 bytes; for Modbus, x, y $>$ 85 bytes. Note that, irrespective of the protocol, the size of the first segment is always 84 bytes.
%\karthik{Sorry, but I don't think this is coming across. Do we really need to explain the packet sequences in so much detail here?}
% \aastha{Agree with Karthik. The text description is anyways difficult to follow; it would be better to have a time diagram to clarify this. If there are space constraints, the diagram can be pushed to the appendix.}
% \gargi{Simplified. Check now.}
% \aastha{A figure will really make things clearer!}

\subsection{Threat Model and Assumptions}\label{sec:threat-model}

\textbf{Adversary position and capabilities.} We consider an adversary with covert, persistent control of one or more ISP or provider-edge routers on the path between the plant gateway and the cloud-hosted SCADA. From there, the adversary can (i) observe traffic metadata such as unencrypted VPN packet headers (IP/TCP), packet lengths, and timings; (ii) record flows for offline analysis; and (iii) drop or delay packets. However, the adversary cannot break encryption, read or inject payloads, or compromise any ICS-internal device.
% \AM{reading payload is same as breaking encryption, no? We are duplicating the point.}

% \hl{Examples of targeted adversaries include rogue ISPs.} %\karthik{Can we relate this to what we said in the intro.} 

Adversaries have compromised ISP or provider-edge routers %is both feasible and 
in the past by exploiting zero-day or unpatched vulnerabilities, misconfigurations, or leaked credentials~\cite{junipervulnerability}. For instance, recently disclosed vulnerabilities in Cisco and Juniper routers~\cite{cisco-1,cisco-2,juniper} grant root access and code execution privileges to remote adversaries, enabling them to monitor traffic using tools such as Wireshark, as well as drop packets. 
This adversary might be a rogue ISP or an enemy nation-state actor who leverages background knowledge to compromise routers known to carry ICS traffic.

Once in position, the adversary isolates the VPN flow carrying ICS traffic. This is straightforward when the endpoint addresses of the VPN tunnel are known: in OT networks, these addresses are long-lived, and a site-to-site tunnel is dedicated to plant-to-SCADA communication, isolated from IT and other-organization traffic for security and performance concerns (NIST SP 800-82)~\cite{nistguidelines}. Otherwise, the adversary proceeds in two steps: separating VPN from non-VPN flows, then ICS from non-ICS VPN flows. The first step is straightforward if the VPN tunnel uses default ports (e.g., 1194/443 for OpenVPN). When it does not, the adversary applies known fingerprinting techniques~\cite{xue2025openvpn, razooqi2025vpn} that rely only on packet direction, size, and ordering, all visible at a compromised router. In OpenVPN, for example, the control messages exchanged during connection bootstrapping or session key renewal produce distinctive packet-size sequences~\cite{xue2025openvpn}. 
% even a correct configuration does not obscure them.
Our testbeds described in \S\ref{subsec:emulated-swat} use tls-crypt-v2, which leaves these fingerprints intact. For the second step, the adversary identifies the ICS flow among the isolated VPN flows by its balanced bidirectional traffic, stable timing, periodic bursts, and long-lived endpoints. Since OT tunnels carry no other organizational traffic, these features persist despite VPN tunneling.

\textbf{Adversary's background knowledge.} 
% We consider a \textit{process-unaware} external (PU) adversary, which is both common and realistic. 
The adversary knows nothing of the target's system design, control algorithm, or implementation. It \textit{knows only the general properties of ICSes}, e.g., that communications are repetitive and periodic, and that sub-processes perform a fixed set of actions.
% The adversary has {no knowledge about the system design, control algorithm, or the implementation details} of the target ICS. However, the adversary is {aware of the general properties of ICS} systems, e.g., the repetitive and periodic nature of ICS operations, and that ICS sub-processes can perform a fixed set of actions. %

% For comparison, we consider two external adversaries on the two extreme ends of the spectrum~\cite{nistguidelines}.
% First, we consider a process-aware external (PA) adversary. 
% This covers adversaries with unauthorized access to the operational manual of the targeted ICS, and rogue ex-employees who exploit their knowledge of the process to attack the ICS. Such attacks have been carried out on water treatment plants in the wild~\cite{maroochy, Ellsworth}.
% However, these adversaries do \emph{NOT} have access to system passwords and traffic decryption keys.
% Second, we consider an external adversary with no general knowledge about ICSes (NK), who can only perform a volumetric DoS attack. Table~\ref{tab:adversary} shows the capabilities of the PU model in comparison with these two additional external adversarial models~\cite{nistguidelines}.
%\karthik{Do we use these later? If not, perhaps we can remove this paragraph and the table, as I think it causes confusion.}\gargi{Removed comparison with other adversary models.}
% \subsection{Adversary Position and Capabilities}

\subsection{Scope}

\sniper{} targets batch-processing ICSes that exchange regular, periodic (synchronous) updates with the SCADA. These systems constitute a large class of safety-critical systems spanning water and wastewater treatment and discrete manufacturing~\cite{batch-water, batch-processing}. We focus on synchronous communication because, when the site updates the SCADA continuously and only a small fraction of packets are critical, it is difficult to separate the critical and non-critical packets. This is the core problem \sniper{} solves. Asynchronous communication, where the site updates the SCADA only on a state change, makes this trivial; so an adversary would not need \sniper{}. 

%% file: sections/methodology.tex
% \newpage
\section{ICS-Sniper}\label{sec:methodology}
We start with a high-level overview of \sniper{} (\S\ref{sec:hlov}), followed by the details of its two phases of operation
% the design challenges, and how we address them using our ETA techniques
(\S\ref{sec:profiling}--\ref{sec:activeattack}).
Finally, we include a discussion of the generalizability of \sniper{} (\S\ref{sec:generalizability}). 
We discuss in Appendix~\ref{sec:comparison} how \sniper{} is better than known volumetric DoS attacks. %in terms of precision and detectability.
% \aastha{Provide forward references to the appropriate subsections.}
%%%%%%%%%% ICS example %%%%%%%%%%%%%%%%%%%%

%%%%%%%%%%%%%%%%%%%%%%%%%%%%%%%%%%%%%%%%%%%%%%%%%%

\subsection{High-Level Overview of \sniper{}}\label{sec:hlov}
% In this section we first present the core idea behind \sniper{}. We then describe the traffic measurements needed to implement the attack, and discuss corner cases where idea would likely fail or be impractical. 
\begin{figure}[t]
    \begin{center}
    \includegraphics[width=\columnwidth*1]{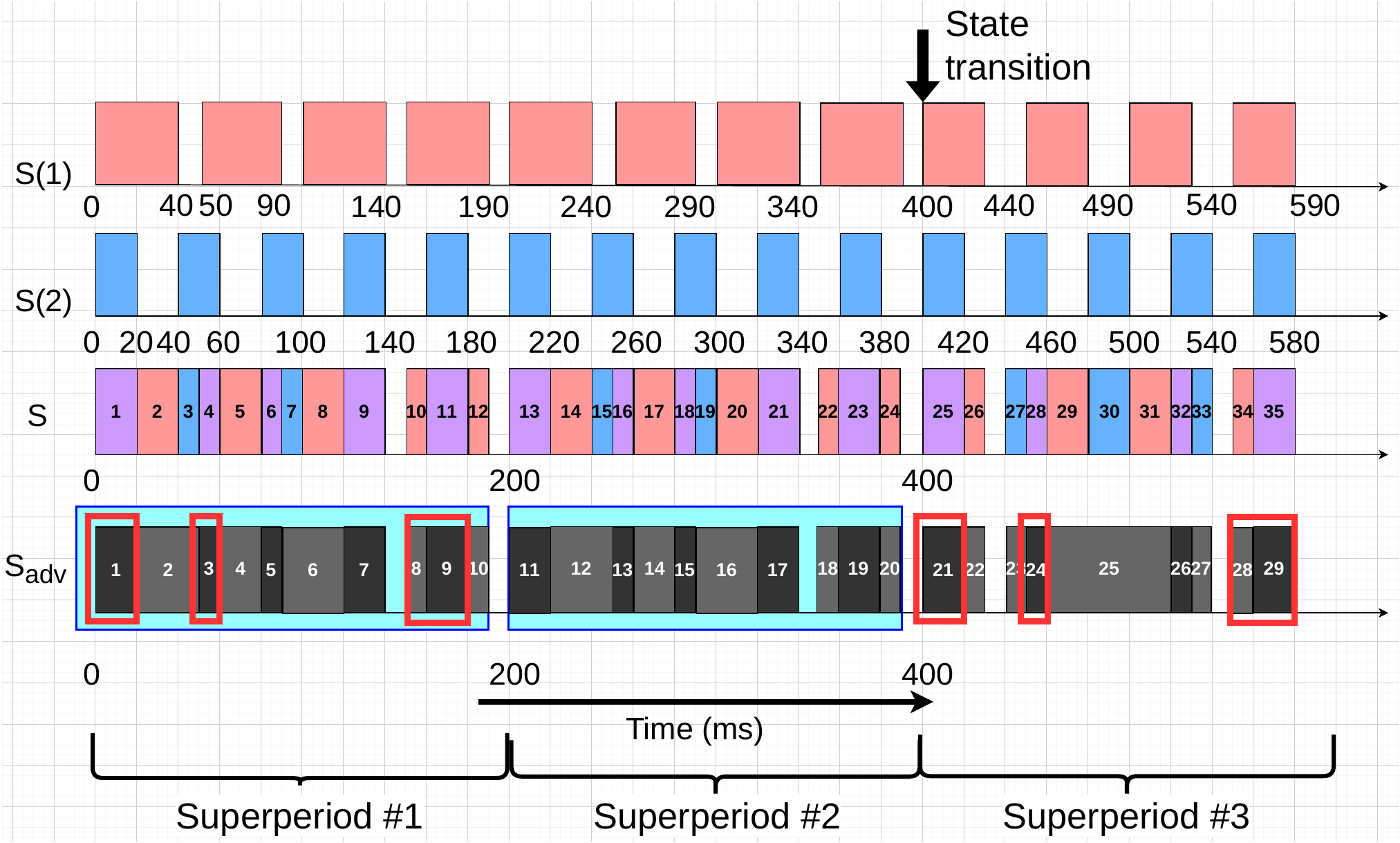}
    % \captionsetup{justification=centering}
    \caption{Packet burst patterns for our example ICS. S(1) and S(2) represent bursts from $plc_1$, and $plc_2$, respectively. S denotes the interleaved burst pattern within the VPN tunnel, and $S_{adv}$ is the adversary's view of the VPN flow. 
    }
    \label{fig:superperiods}
    \end{center}
\end{figure}
% \aastha{High-level overview of what?}
% \aastha{Explain in the caption of Fig 3 itself, what S(1), S(2), and S are.}\gargi{Done}

% \subsubsection{Attack strategy}\label{sec:core-idea}
% \aastha{Intuition for what?}
\textbf{Premise.} As discussed in \S\ref{subsec:ics-state-trans}, each PLC generates a periodic bursty packet flow throughout an operational cycle, with each burst carrying data requests for remote sub-processes and its own process updates to the SCADA. The number and sizes of packets within a burst characterize the PLC's current state, and a change in burst pattern signals a state transition.
%In the absence of 
Without knowing the plant's control logic, \sniper{} operates on the intuition that \textit{a change in a burst pattern indicates a state transition in the corresponding sub-process, triggered by one or more packets in either the transitioning burst or the one immediately preceding it.} \sniper{}'s goal is to identify and drop these bursts to disrupt cross-sub-process synchronization. 
% Since our adversary has no knowledge of the plant's control logic, it operates on the intuition that \textit{a change in a PLC's burst pattern indicates a state transition in the corresponding sub-process, triggered by one or more packets in either the burst with the new pattern or the burst immediately preceding it.} Therefore, the goal of the adversary is to identify these packet bursts and drop them to disrupt cross-sub-process synchronization.  

\textbf{Main challenge.} In an ICS, multiple gateway PLCs might share a single VPN tunnel to the cloud-hosted SCADA, aggregating their packet streams into one encrypted flow. Because VPN encapsulation hides the per-PLC IP/TCP headers, it is challenging for the adversary to discern per-PLC traffic based solely on observable metadata of this aggregate flow. %\karthik{Number the challenges and refer to them in the text}\gargi{In high-level overview, we have mentioned only the main challenge. In the descriptive subsections, I have numbered the challenges and referred to them in text.}

% A change in burst pattern in period $p_i$ implies that the state transition is either driven by the output computed in the $p_{i-1}^{th}$ period
% % \aastha{Remove the phrase "based on the inputs collected in the $p_{i-1}^{th}$ period"?}\gargi{Done},
% or the inputs collected in the $p_{i}^{th}$ period. The SCADA must deliver the packets carrying these inputs to the PLC within a bounded time, i.e., one PLC period, to preserve inter-sub-process synchronization. Our adversary aims to detect a change in burst pattern between consecutive periods and drops all packets in the period where the change appears and in the immediately preceding period.

\textbf{Core Idea.} To address this challenge, \sniper{} exploits a key insight: although VPN encryption obscures per-PLC traffic patterns, the periodic bursts of individual PLCs produce a structured aggregate flow characterized by alternating regions of overlapping (concurrent PLC activity) and non-overlapping (single or no active PLC) bursts. This structure gives rise to a periodic pattern in the aggregate flow, which we call \textit{superperiodicity}. The sequence of overlap and non-overlap regions recurs with a fixed period, the \textit{superperiod} L, equal to the least common multiple of the individual PLC communication periods. While \sniper{} cannot identify which specific sub-process undergoes a state transition, it can identify the superperiods during which one or more sub-processes do, and use this to isolate a small subset of packets critical to cross-sub-process synchronization, and selectively drop them.

\textbf{Illustration.} %We illustrate this concept using our example ICS. 
Figure \ref{fig:superperiods} shows the traffic burst patterns of the two PLCs introduced in Figure \ref{fig:st-example}:
S(1) for $plc_1$ (period 50 ms, initial burst length 40 ms) and S(2) for $plc_2$ (period 40 ms, burst length 20 ms).
% Initially, S(1) has a burst length of 40ms, and S(2) has a burst length of 20ms. 
We assume that $plc_1$ transitions from State-1 to State-2 after eight periods (i.e., after 400 ms), reducing its burst length from 40 ms to 30 ms as it transmits fewer packets in State-2. The burst duration of S(2) remains unchanged since $plc_2$ exchanges the same set of messages with the SCADA in both states. S represents the aggregate traffic resulting from the multiplexing of the periodic bursts S(1) and S(2). The purple regions indicate time intervals where bursts from both PLCs overlap, causing their packets to appear interleaved. The sequence of overlap and non-overlap regions recurs periodically, forming the superperiodic structure of S.
% \aastha{Visually, it is hard to identify the superperiodic structure of S. Can you show two superperiods in the diagram that truly identical; this is only for the sake of introducing the concept. Also, the braces at the bottom of the figure showing superperiod 1 and 2 are hardly visible.}\gargi{Modified the figure to incorporate the suggestion.}
Overlap regions 1-12 in S repeat in regions 13-24 with a period of 200 ms. Therefore, the superperiod of the aggregate traffic S is 200 ms (= LCM (50 ms, 40 ms)). After 400 ms, although a change in a PLC's burst pattern may alter the appearance of some overlap regions, other regions continue to exhibit the same superperiodicity.
In Figure \ref{fig:superperiods}, despite the change in S(1) during its ninth period, regions 1, 4, 5, 8, 10, and 11 recur in regions 25, 28, 29, 31, 34, and 35 with a period of 200 ms. Therefore, even if state transitions alter the traffic patterns of some overlap regions within a superperiod, the remaining recurring overlap regions aid in determining the superperiod.
% \aastha{Point of the last two sentences not clear.}

% \gargi{Repeating}Since the VPN encapsulation hides the per-PLC IP/TCP headers, it is challenging for the adversary to associate packets with their respective PLCs based solely on observable metadata. 
Similar to S, its encrypted version $S_{adv}$ also consists of alternating overlap and non-overlap regions. However, in the absence of per-PLC identifiers, $S_{adv}$ lacks certain structural periodic information that is otherwise visible in S. For example, regions 2 and 3 in S appear as one contiguous non-overlap region (region 2) in $S_{adv}$. Nevertheless, even under VPN encryption, portions of the traffic in $S_{adv}$ still exhibit superperiodic recurrence. Overlap regions 1-10 in $S_{adv}$ repeat in regions 11-20 with a periodicity of 200 ms (highlighted regions), exhibiting a superperiod of 200 ms (= LCM (50 ms, 40 ms)). Even after the state transition of $plc_1$, regions 1, 3, 8, and 9 in S reappear as regions 21, 24, 28, and 29, respectively, exhibiting a period of 200 ms.

\textbf{ETA technique and associated challenges:}
\sniper{} leverages traffic metadata to identify recurring overlap regions and infer the underlying superperiodic structure and burst pattern changes.
% \textcolor{blue}{
However, network noise and frequent state transitions can distort the overlap regions, making superperiod estimation challenging.
% Specifically, it leverages three metadata features (referred to as events): unique packet sizes, inter-packet interval durations, overlap region durations.
% Network noise and frequent state transitions can introduce variance in these features, distorting the periodic structures and ultimately making superperiod estimation challenging.
% distort the periodic structures, making superperiod estimation challenging.
% }
% This is challenging because network-level noise and frequent state transitions can distort superperiodic structures.
% 
\sniper{} addresses this challenge using phase folding (described in \S\ref{sec:phasefolding}), a noise-resilient technique for detecting periodic structures.
% While overlap region identification uses inter-packet intervals, \sniper{} also independently applies phase folding to unique packet lengths and interval occurrences as discrete events, providing complementary superperiod estimates that do not rely on overlap regions.
% 
% \textcolor{blue}{
Specifically, it applies phase folding to three traffic metadata features (referred to as events): overlap region durations, inter-packet interval durations, and unique packet lengths. We refer to the three steps as $\phi_{overlap}$, $\phi_{interval}$ and $\phi_{pktlen}$, respectively.
First, in the $\phi_{overlap}$ step, {\sniper} applies phase folding to the coarse-grained feature overlap-region durations, which yields the superperiod value. However, noise in the overlap region durations can introduce errors in the estimated superperiod value (e.g., yielding a superperiod of 179.5s instead of 180s), which can accrue over time and significantly misalign the inferred boundaries of later superperiods in the operational cycle.
Therefore, {\sniper} also applies phase folding to the finer-grained features, inter-packet timing intervals (IPTs) and unique packet lengths in $\phi_{interval}$ and $\phi_{pktlen}$ steps, respectively. These steps yield precise period values of one or more PLCs (e.g., 60s, 90s, etc.).
{\sniper} determines the correct superperiod value as the least common multiple of the candidate periods generated in $\phi_{interval}$ and $\phi_{pktlen}$  that is closest to the superperiod estimated from 
$\phi_{overlap}$.
% Phase folding is noise-resilient because it can recover the underlying superperiod even when network noise and state transitions cause some overlap regions to disappear or shift slightly.
% It does not require all overlap regions to align perfectly across every superperiod.
% However, noise can introduce small variations in overlap durations, which although are too minor to be treated as distinct overlap regions, are still large enough to introduce errors in superperiod estimation (e.g., yielding a superperiod of 179.5s instead of 180s).
% For a targeted attack like {\sniper}, these small errors in superperiod estimation can accrue over time and significantly misalign the inferred boundaries of later superperiods in the operational cycle.
% }

\textcolor{blue}{
}

\textbf{Stages:} \sniper{} operates in two stages.
% \aastha{We overload the term phase several times. It is mentioned here, then is part of the phase folding technique. Within the profiling phase, you again have mid-operational and idle phase. Check and disambiguate. Here replace phase with stage.}
In the \textit{Process Profiling Stage}, it records OT traffic from a compromised router and analyzes traffic metadata to infer the superperiod length L, and subsequently detects state transitions. In the \textit{Active Attack Stage}, it monitors live traffic and drops critical packets at predicted times. If \sniper{} detects a change in burst pattern within the $x^{th}$ superperiod, it drops packets in both the $(x-1)^{th}$ and $x^{th}$ superperiods.

\subsection{Process Profiling Stage}\label{sec:profiling}
% We set up a dummy VPN-connected PLC–SCADA pair, where the PLC was hosted on a local machine and the SCADA was hosted on a cloud server, and exchanged random messages between them to uncover the characteristics of VPN-encrypted OT traffic.
%\aastha{Comes out of the blue. Which set up are we talking about? Doesn't the testbed get described later?} \karthik{I agree. Also, what do you mean by random? Did you sample the communication at random or did they exchange random messages?}\gargi{Moved this and clarified in Section 3.2}

% Now, \sniper{} leverages these observations to identify the superperiods to target, from the combined VPN traffic. It does so in the following steps.

\smallskip\textbf{Step 1: Capture the network traffic for a complete operational cycle of the target sub-process.}
\sniper{} must record all VPN packets spanning a complete operational cycle, as missing packets can compromise accuracy. We discuss two challenges in this step and how \sniper{} addresses them.

\textit{C1: Identifying the first and last packets of a cycle.
% from encrypted traffic without payload visibility or access to the plant manual.
}
% 
% \textit{Solution.}
Batch-processing ICSes alternate between an active operational phase (high network activity) and an idle loading phase (negligible activity). In the example from \S\ref{sec:prelims}, the plant operates for 10 hours followed by a 14-hour idle period. IPTs differ significantly between these phases, allowing \sniper{} to detect phase boundaries. If \sniper{} begins capturing at an arbitrary point, it would encounter one of the two scenarios. \textbf{(1) Mid-operational phase.} 
% \sniper{} monitors inter-packet intervals and, upon observing one that significantly exceeds the running average, infers that the current cycle has ended. It discards all prior packets, treats the next arriving packet as the start of a new cycle, and monitors until another large interval signals its end.
On observing an IPT that far exceeds the running average, \sniper{} infers the current cycle has ended, discards all prior packets, treats the next packet as a new cycle's start, and waits for the next large interval to mark its end.
\textbf{(2) Idle phase.} \sniper{} detects an extended inactivity period followed by packets with low IPTs, and captures all subsequent packets until the next inactivity period signals the cycle's end.

\textit{C2: Handling packet variance due to natural network fluctuations and eliminating non-PLC packets}.
% 
% \textit{Solution.}
Using TCP sequence numbers, \sniper{} removes retransmitted VPN packets, and by parsing VPN headers (see \textbf{O1}) it extracts encrypted packet sizes, discarding any outside the protocol-specific ranges (see \textbf{O2}). It likewise discards heartbeat, keep-alive, and liveness-poll packets so they do not perturb the superperiod estimate; their single-byte payloads result in fixed sizes outside the protocol-specific ranges (O2), making them easy to identify. This is repeated over multiple cycles until a consistent packet count recurs for the majority, after which one representative sequence is retained for subsequent processing.

% \sniper{} removes retransmitted VPN packets using TCP sequence numbers and extracts encrypted packet sizes by parsing VPN headers (see \textbf{O1}), discarding packets with sizes outside protocol-specific ranges (see \textbf{O2}). \gargi{Point\#2, Change\#2}Similarly, \sniper{} \hl{also discards heartbeat, keep-alive, and liveness-poll packets so that they do not perturb the superperiod estimation. Their single-byte
% payloads make them fixed in size and place them outside the protocol-specific size ranges (see \textbf{O2}), which makes them straightforward to identify.} This process is repeated across multiple cycles until a consistent packet count is observed for the majority of cycles, after which one representative sequence is selected for subsequent processing.

\textbf{Step 2: Metadata extraction.}
% In this step, \sniper{} extracts the metadata of the VPN packet sequence captured in Step 1.
For each packet in the captured trace, \sniper{} extracts the source and destination IP addresses, its arrival time relative to the beginning of the operational cycle, and the sequence of encrypted packet sizes (similar to Step 1), yielding a sequence S of tuples ($l_t,t$). %Here, 
$l_t$ is a signed integer representing an encrypted packet length (positive for PLC $\rightarrow$ SCADA and negative for SCADA $\rightarrow$ PLC) and $t$ is the timestamp relative to the first packet in the operational cycle. Encrypted packets that were coalesced into the same VPN packet are assigned the same timestamp $t$. 
For determining which IP address belongs to the PLC and which one belongs to the SCADA, \sniper{} applies \textbf{O2} in \S\ref{sec:vpnchar} for unconnected variants, i.e., IP addresses generating 60\% of encrypted packets must belong to the site gateway, while IP addresses generating 40\% of encrypted packets must belong to the SCADA. For connected variants, the IP address assignment is interchangeable.

\begin{figure}[t]
    \begin{center}
    \includegraphics[width=\columnwidth*1]{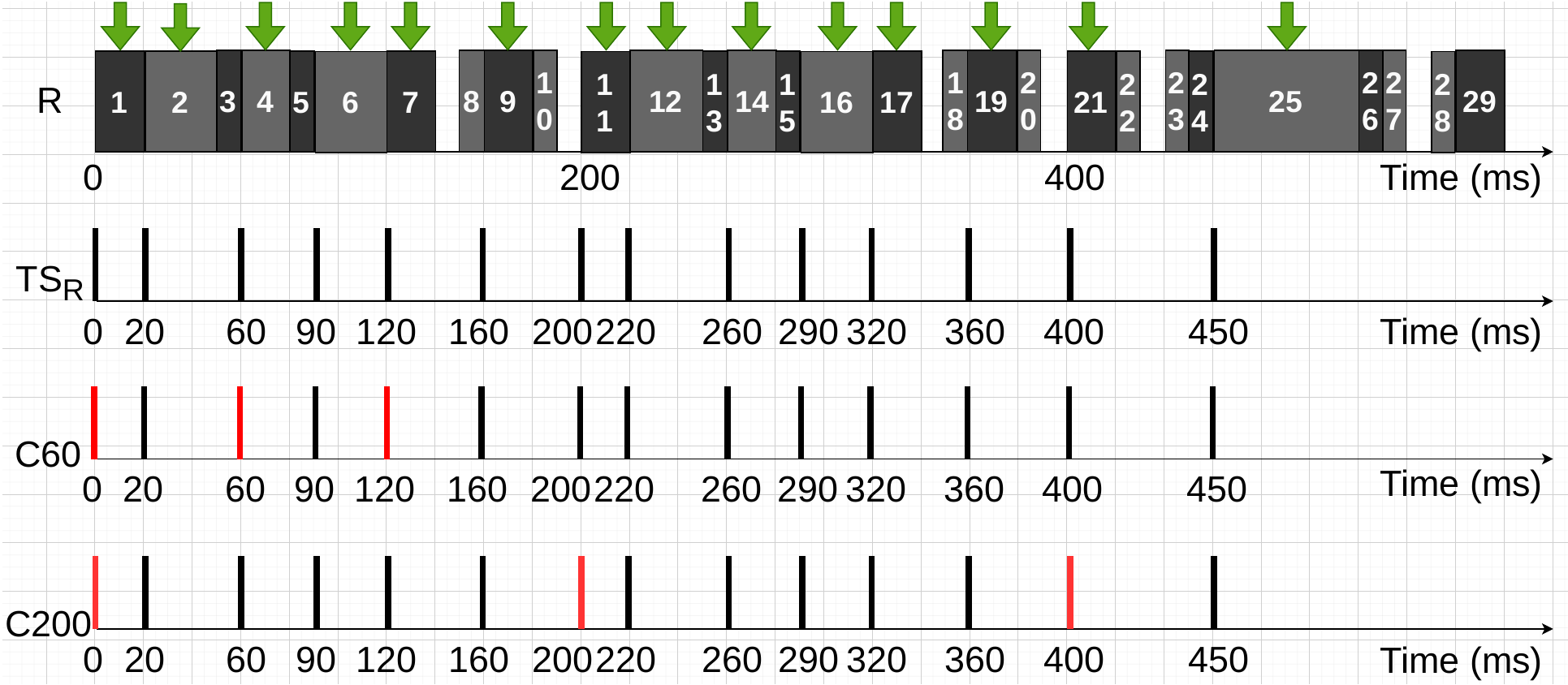}
    % \captionsetup{justification=centering}
    \caption{Identifying stable regions and superperiods from overlap regions. The regions marked with green arrows indicate stable regions. $TS_R$ is the resultant time series. The red lines in $C_{60}$ and $C_{200}$ show events that line up while phase folding with 60ms and 200ms candidates respectively. 
    }
    \label{fig:stable-reg}
    \end{center}
\end{figure}

\textbf{Step 3: Identifying the superperiod L.}
% Given the sequence $S$ from Step 2, \sniper{}'s goal in this step is to identify the superperiod $L$.
% To this end,
% \aastha{Inconsistent formatting of symbols. Sometimes they are in math mode and sometimes in regular mode. Use macros to define a symbol and present them with a consistent formatting.}
To identify the superperiod L, \sniper{} processes the sequence S from Step~2 using phase folding~\cite{zucker2015detection} (described in \S\ref{sec:phasefolding}). 
%Now 
We discuss how \sniper{} applies this technique while considering the three traffic metadata features as events: overlap region durations, IPT durations, and unique packet lengths. For each event type, \sniper{} constructs two event series, based on traffic direction (PLC $\rightarrow$ SCADA and SCADA $\rightarrow$ PLC). 

\textbf{Step $\phi_{overlap}$}: \sniper{} delineates overlap regions from shifts in IPTs. 
From S, it computes the IPT between each pair of consecutive packets, yielding a sequence S'. It then slides a 2-element window over S', computing the absolute difference between each consecutive IPT pair. If this difference exceeds a threshold, a region boundary is placed between the two packets corresponding to the larger IPT difference. This boundary indicates a transition between an overlap and a non-overlap region, or between two overlap regions with differing numbers of concurrently active PLCs. However, this technique cannot distinguish between two non-overlap regions belonging to different PLCs, since all PLCs transmit messages at similar rates. The threshold is determined via hierarchical clustering on all consecutive pairwise IPT differences in S', which partitions them into maximally separated groups. \sniper{} then selects the largest value in the smallest-difference cluster as the threshold, i.e., the maximum IPT variation for packets belonging to a single region. This yields a sequence of regions R=\{($r_{start}$, $r_{end}$)\}, where $r_{start}$ and $r_{end}$ are the start and end times of a region.

In our example shown in Figure~\ref{fig:superperiods}, this step produces 29 regions for our ICS example continued from \S\ref{sec:hlov}. Of these regions, \sniper{} uses the longest ones for periodicity analysis, as they indicate stable regions. For each region in R, \sniper{} computes its duration as ($r_{end}$-$r_{start}$) and applies hierarchical clustering to the resulting set of durations to filter the long-duration regions that serve as reliable reference points for period estimation. Let us assume, in our example, all regions of duration $\ge 20$ms are considered stable. This produces 14 stable regions, indicated by the green arrows in Figure~\ref{fig:stable-reg}. 
From their start times, \sniper{} builds a time series ($TS_R$ in Figure~\ref{fig:stable-reg}), yielding 36 candidate periods. Since at least 3 events must align for a superperiodic structure, candidates exceeding one-third of the cycle length (200 ms here) are eliminated, leaving 18. Figure~\ref{fig:stable-reg} shows phase folding for two candidates, 60 ms and 200 ms (red lines mark aligned events): 3 events align for both, but 200 ms spans more of the operational cycle. Evaluating the remaining candidates likewise identifies 200 ms as the superperiod in this example.

% Once the stable regions are identified, \sniper{} constructs a time-series with the start times of the stable regions ($TS_R$ in Figure~\ref{fig:stable-reg}). This produces 36 candidate periods, but since at least 3 events have to align for a superperiodic structure, we can eliminate all candidates that exceed one-third of the operational cycle length (200ms in this example). This leaves us with 18 candidates. Figure~\ref{fig:stable-reg} shows the result of phase folding for two such candidates -- 60 ms and 200ms. The red lines indicate events that line up when the event timeline is folded into strips of these lengths. In this example, for both candidates, 3 events lined up. However, the 200ms candidate spanned more of the operational cycle than the other candidate. Evaluating all other candidates also revealed 200ms to be the superperiod in this case.

\textbf{Step $\phi_{interval}$}: Short PLC communication periods combined with network jitter can make stable region identification unreliable. \sniper{} therefore also phase-folds the IPTs directly, treating each as an event. 
It clusters the IPT values in S' by hierarchical clustering—maximizing separation while grouping similar intervals—so that each cluster captures the IPTs arising when a given number of PLC periods coincide, then builds a time series per cluster and phase-folds it as in the previous step.
% It first groups IPT values in S' into clusters using hierarchical clustering, maximizing separation between clusters while keeping similar intervals together. Each cluster captures IPTs that arise when the same number of PLC periods coincide. A time series is then constructed for each resulting cluster. Once the time-series is constructed, the phase folding technique is applied in a similar manner as the previous step. 
The period determined for each cluster would either coincide with the period of a PLC or the Least Common Multiple (LCM) of the periods of a subset of PLCs.

\textbf{Step $\phi_{pktlen}$}: 
\sniper{} exploits the payload size variability of OT packets (\textbf{O3} in \S\ref{sec:vpnchar}). If a PLC has at least one packet size recurring across multiple communication cycles, phase folding can recover its communication period. When this holds for all PLCs, the LCM of the recovered periods results in L. Otherwise, only a subset of periods is recovered. Because \sniper{} lacks ground truth, it always runs all three event types. Even partial length-based estimates help refine the coarse overlap-based estimate. \sniper{} identifies the unique packet sizes in S and, for each, builds a time series using the timestamps at which that size appears and phase-folds it. The recovered period matches the period of one or more PLCs transmitting packets of that length.

\textbf{Step 4: Predicting state transitions.} Once \sniper{} determines the superperiod L, it splits the metadata sequence S into sub-sequences of duration L and compares each adjacent pair of superperiods using two heuristics, each capturing a different aspect of traffic variation:
% (i) number of overlap regions that differ in duration, which reflects changes in PLC burst durations,
(i) change in total payload packet count, which captures changes in the number of packets due to state transitions, and
(ii) change in the number of unique packet sizes, indicating new message sequences due to state transitions.
Individually, each heuristic is imperfect: network jitter and retransmissions can affect packet counts, while spurious control packets can introduce unexpected message sizes. However, a combination of heuristics serves as a strong indicator of a state transition.
\sniper{} designates a superperiod as \textit{critical} if either its total packet count or its number of unique message sizes, or both, differ from the corresponding values in the preceding superperiod. If multiple superperiods are designated critical, \sniper{} ranks them
by first counting how many heuristics show a difference (one or both), and prioritizes the one with differences across the maximum number of heuristics. A tie is resolved by the combined magnitude of those differences, computed as the sum of the absolute differences in packet count and unique message sizes relative to the preceding superperiod. The superperiod with the highest-ranked discrepancy is assigned the highest priority.
\textbf{Corner case analysis.}
% \aastha{Why are these subsubsections? What are they subsubsections of? Step 4? There is no sentence explaining the relation of these to the previous paragraph. I would put them at the same level as the four steps because I think they are not related to step 4 alone but rather the whole process profiling phase.}
\sniper{} is likely to mispredict state transitions or fail to identify the superperiod under the following unusual yet theoretically feasible scenarios: transitions driven purely by local sub-process changes; concurrent transitions that leave aggregate traffic metadata unchanged; very frequent per-PLC transitions that obscure superperiod patterns; or a superperiod (the LCM of PLC periods) that is too large to repeat within the operational cycle. However, this is highly unlikely in practice, since real PLC periods range from a few milliseconds to seconds, while operational cycles span hours. Appendix \ref{subsec:corner-case} discusses these scenarios in detail.
% \aastha{Reference the appendix section number here.}
%\gargi{Moved the details to the Appendix, as suggested}

% \subsubsection{Attack Success vs. Detectability Trade-off}
% If the PLC periods share many mutually prime factors, the resulting superperiod becomes longer. While a longer superperiod can encompass multiple critical state transitions, potentially increasing \sniper{}'s success probability, it also increases the likelihood of compromising on stealth.
% \aastha{So what? Why are you telling me this? What is a reader supposed to do with this information? This is an abrupt ending.}
%%%%%%%%%%%%%%%%%%%%%%%%%%%%%%%%%%%%%%%%%%
\subsection{Active Attack Stage}\label{sec:activeattack}
% \gargi{Heuristics for prioritizing critical superperiods are already mentioned in Step 4 of profiling. So not repeating it here.}
In this stage, \sniper{} drops the packets in the critical superperiods in real-time. 
It observes the traffic and waits to detect the start of a new operational cycle, identifying the first packet with the technique from Step 1 of the profiling stage (\S\ref{sec:profiling}). If the target is the $x^{th}$ superperiod in S, \sniper{} drops payload-carrying packets in the $(x-1)^{th}$ and $x^{th}$ superperiods, i.e., for an operational cycle starting at time $T_s$, within $[(T_s + (x-2)\times L), (T_s + xL)]$. Only application-payload packets are dropped: 20\% of the packets in that window for unconnected protocol variants (\textbf{O2} in \S\ref{sec:vpnchar}), and all packets in the connected variant, where every packet carries payload. Dropping payload packets across this window also drops all retransmissions, effectively preventing the %scheduled 
state transitions.
Appendix~\ref{sec:comparison} compares this to volumetric DoS techniques. 
%\karthik{Moved the volumetric comparison to the appendix}

\subsection{Generalizability}
\label{sec:generalizability}

\textbf{Across protocols.} \sniper{} depends only on the periodic request-response structure of ICS communication, not on any protocol-specific field. Any protocol that follows a request-response pattern with periodic updates, e.g., MMS in IEC~61850~\cite{mms}, produces the same superperiodic, balanced traffic pattern that \sniper{} exploits. This pattern is also preserved when several similar protocols share a flow (e.g., Modbus + MMS, as in the EPIC power-grid dataset~\cite{epic}).

\textbf{Across VPNs.} \sniper{} can be adapted to different VPNs with no changes. Every VPN protocol that secures site-to-SCADA traffic encrypts each inner 
IP packet 1:1, preserving the timing, ordering, and burst structure that \sniper{} relies on. Handshake and key-exchange packets, which do not carry process data, are identified and discarded by their characteristic fixed sizes so that they do
not perturb the superperiod estimate. The only factor that might affect \sniper{}'s profiling accuracy is ciphers that coarsen packet size and thus reduce size resolution. In this case, \sniper{} falls back on timing alone, which we show is sufficient to recover the superperiodic
structure~(\S\ref{subsec:eval-lts}).
%%%%%%%%%%%%%%%%%%%%%%%%%%%%%%%%%%%%%%%%%%%%%%%%%%%%%%%

%%%%%%%%%%%%%%%%%%%%%%%%%%%%%%%%%%%%%%%%%%%%%%%%%%%%%%%
% \subsection{Encrypted Traffic Analysis Techniques}

%% file: sections/experimental-setup.tex
\section{Experimental Setup}\label{sec:experimental-setup}
We first describe the original Secure Water Treatment (SWaT) plant setup at a high level~\cite{swat-testbed}.
% and the original testbed simulating this setup~\cite{swat-testbed} in a single process.
We then describe our enhanced testbed that emulates a plant with SCADA deployed on the cloud. We then describe the implementation of the attacks in our setup and also the attack detection techniques that we later use to evaluate {\sniper}'s efficacy. 

\subsection{SWaT Plant}
\label{subsec:swat-plant}
% Figure~\ref{fig:testbed} in Appendix~\ref{sec:hlov-appendix} shows the architecture of the complete SWaT plant~\cite{swat-testbed}.
We evaluated ICS-Sniper on testbeds modeled after the miniature 6-stage Secure Water Treatment (SWaT) plant at SUTD~\cite{swat-testbed}. To our knowledge, this is the only openly available realistic multi-sub-process control logic, allowing us to evaluate \sniper{}'s impact on the plant's operations. 
The plant purifies wastewater in six stages, each controlled by a sub-process: raw water storage ($P1$), chemical dosing ($P2$), ultrafiltration ($P3$), dechlorination ($P4$), reverse osmosis ($P5$), and backwash ($P6$).
The testbed consists of six PLCs, each acting as the gateway of a sub-process. The number of internal PLCs behind each gateway does not affect our evaluation, as %. Internal PLCs do not communicate with the SCADA, and} 
\sniper{} targets only gateway-PLC $\rightarrow$ SCADA traffic. 

SWaT itself has six PLCs, all of which
are gateways reporting to the SCADA. Where a real plant would place additional PLCs behind a gateway, SWaT represents that internal layer by a control logic, which acts on a simulated model of the mechanical components it governs. The resulting component actions drive the overall physical process, which is itself mathematically modeled. To maintain synchronization across all six sub-processes, the PLCs periodically exchange information about the current state of their physical components and other sub-processes with the SCADA system. The PLCs communicate with the SCADA over the Intranet using unencrypted communication channels.
% \textcolor{blue}{
We provide additional description of the original SWaT plant in Appendix~\ref{subsec:swat-supplementary}.

\subsection{Internet-connected SWaT Testbed Emulation}\label{subsec:emulated-swat}
% \textbf{The original SWaT testbed.} This
% However, this testbed does not accurately represent the target environment of ICS-Sniper because of two reasons – (1) this testbed is not connected to the Internet. The PLCs communicate over Intranet; and, (2) this testbed does not use encrypted inter-PLC communication protocols. Therefore, we built a testbed that mimics the operations of the SWaT testbed, but in a geo-distributed environment, and uses encrypted communications.
% Our emulated testbed mimics a geo-distributed water treatment plant where the first stage (raw water collection) is
% segregated
% % located geographically apart
% from the other five stages (water treatment) in two separate ``sites'' (referred to as site A and site B). After collection, the raw water is transmitted from the first site to the second via underground pipelines.
% The PLCs of site A and site B communicate over network using encrypted protocols.
% The PLC in the first stage communicates with the other PLCs over the Internet using encrypted protocols. 
% 
We built two testbeds that mimic a geo-distributed version of the SWaT plant, where the synchronization among its six sub-processes is maintained by a central cloud-hosted SCADA system that communicates with the PLCs over the Internet via an OpenVPN tunnel. Both testbeds allow configuring different PLC periods and simulating various network conditions.

\textbf{(i) S-SWaT:}
% \textit{(a) Process simulation.} 
Our first testbed contains software-simulated PLCs to enable fast prototyping. Unlike the original physical testbed~\cite{swat-testbed}, our emulation does not have any real mechanical devices. Instead, we use the control logic and physical process models available from an open-source single-host Python simulation of the same testbed~\cite{swat-simulator}. We implemented the control logic of the 6 sub-processes and the physical process models as separate Python modules, and we execute them as separate processes within an Amazon EC2 VM, which represents the plant site. 
% While the original Python simulation implemented PLC-to-SCADA communications via shared memory, we replaced them with network-based communication.
The traffic of all 6 PLCs is routed through the same network interface of the EC2 VM, resembling the plant site gateway. S-SWaT has an operational cycle of 4 hours. 

% \textit{(b) Network Protocols with Encryption and VPN Support.}
For site-to-SCADA communication, S-SWaT provides configurable support for unconnected variants of two industry-relevant protocols: ENIP/TCP and Modbus/TCP. For securing the communications, we encapsulated the site-to-SCADA traffic inside an OpenVPN tunnel that uses TLS Crypt v2,
% (an improved version of OpenVPN's control channel encryption mechanism that uses client-specific keys)
AES-256-GCM for data encryption, and SHA-256 to verify message integrity. This ensures adequate confidentiality, integrity, and authenticity of OT communications. 

% \textit{(c) Cloud deployment.}
We deploy the testbed in the cloud using three Amazon EC2 VMs (see Figure~\ref{fig:swat-testbed} in Appendix~\ref{subsec:swat-supplementary}). VM-1 simulates the plant (ICS site) and runs the six simulated PLCs. VM-2 runs the SCADA system, which logs sensor readings every 10 seconds for telemetry and attack detection. VM-3 acts as an on-path compromised router, through which we route all OpenVPN packets between the site and the SCADA. 
This setup allows for analyzing the ICS traffic patterns in realistic network settings.

% \textit{(d) Attack execution.}
We consider an adversary who targets the site-SCADA communication.
We place {\sniper} on VM-3 (Figure~\ref{fig:swat-testbed} in Appendix~\ref{subsec:swat-supplementary}) to emulate an adversary on the network path connecting the plant VM gateway to the SCADA VM gateway.
{\sniper} relies on (i)~a 
packet capturing tool (we use tshark~\cite{tshark}), (ii) a process profiling module implemented in Python by us, and (iii) a network controller module for dropping packets, implemented in Python by us.

\textbf{(ii) H-SWaT:} This testbed uses real industry-standard PLCs (Allen-Bradley Micro800 series) for reproducing the timing and communication characteristics of real deployments (validated by comparing with the traffic characteristics of the original SWaT testbed). 
We converted the Python-based control logic to structured text code and deployed it on the real PLCs using the Rockwell Automation Connected Components Workbench software. %\karthik{Mention the software}\gargi{Done}. 
The 6 PLCs are connected to a gateway router that is connected to the Internet. For site-to-SCADA communication, H-SWaT uses connected Modbus/TCP encapsulated in an OpenVPN tunnel (see Figure~\ref{fig:hswat} in Appendix~\ref{subsec:swat-supplementary}). H-SWaT is faster in processing and communication, and has an operational cycle of 2 hours, with the SCADA logging sensor readings every second. The rest of the testbed (physical process model, SCADA, simulated on-path compromised router, and attack execution infrastructure) is the same as S-SWaT. To the best of our knowledge, this is the first validated ICS testbed with SCADA in the cloud. 

\subsection{ICS Attack Detection Techniques}\label{sec:detectors}
% 1. Name of the detector,
% 2. Type of detector (process log based or network trace based, and if network trace based, what all does it look at)
% 3. What kind of attacks have they been tested on previously and why we think it is a good fit for our paper
% 4. What modifications / fine-tuning you had to do to make them work for our attack and why you did them
To our knowledge, there is no technique for effectively detecting targeted blackhole attacks on ICS OT traffic. Existing ICS attack detection techniques~\cite{lin2018iat, mirsky2018kitsune, cheng2019invariant, aoudi2018pasad, wolsing2022simple} were tailored for detecting volumetric DoS attacks from network traffic or anomalous sensor readings from process logs in real time. Nevertheless, we assess their capability to detect attacks performed by {\sniper}. 
% In total, we use 8 detectors: \textbf{(i)} 3 network traffic-based detectors - IaT-mean and IaT-range from ~\cite{lin2018iat}, and Kitsune~\cite{mirsky2018kitsune}, and \textbf{(ii)} 5 process log-based detectors - InvariantRules~\cite{cheng2019invariant}, PASAD~\cite{aoudi2018pasad}, Histogram, MinMax and Steadytime~\cite{wolsing2022simple}.
Additionally, we use IPAL~\cite{wolsing2024ensemble}, an ensemble of these state-of-the-art ICS attack detectors. IPAL uses two relevant ensemble strategies: All and Any~\cite{wolsing2024ensemble} to combine the alerts raised by the individual detectors mentioned above. They improve the False Positive Alarms (FPA) and False Negative Alarms (FNA) of the ensemble, respectively. As these ICS attack detectors are directly relevant to ICS-Sniper, we evaluate against them in \S\ref{subsec:eval-detection}. However, conventional cloud-based attack detectors~\cite{aws-shield, reece2024defending, fayaz2015bohatei} are orthogonal to our threat model (see \S\ref{sec:rel-work}). 
% \karthik{I wonder if this information can be better summarized in a table? It's quite dense}\gargi{Moved to table 3}
% \karthik{I'm confused; how is this different from the "Choice of detectors" above? Is that for the ensemble. Overall, I'm not really sure we need so much detail - perhaps a table is better. }
% The inference times of the selected detectors remain within 15\% of the communication cycle in our testbed, ensuring their capability for real-time intrusion detection. 

% Just like \sniper{}, the network trace-based detectors extract features like timing information, communicating entities, function codes, and data values from network traces to build their detection models.
% Most of the above detectors are designed to detect process value anomalies, while a few are designed to detect network-level anomalies, such as Denial of Service, Man in the Middle, and TCP sequence prediction attacks.
\textbf{Choice of detectors.} The detectors in the ensemble represent a diverse set of detection techniques and have been evaluated in prior work on various ICS datasets, including SWaT~\cite{goh2017swatdataset} and WADI~\cite{ahmed2017wadi}. The network anomaly detectors work by analyzing the content and metadata of unencrypted network traffic captured at the SCADA, and the process anomaly detectors flag anomalies in the sensor and actuator values caused as a result of the \sniper{} attack.
Each of these detectors considers different features to detect anomalies in ICS, thereby covering a broad range of features and techniques. Table~\ref{tab:detectors} summarizes the detectors by their type, technique, and features used for anomaly detection.
% Table~\ref{tab:detectors} summarizes the techniques and features used by the anomaly detectors in the ensemble.

% \textit{DTMC} builds a discrete-time Markov chain model in which states represent system states based on sequences of network messages observed during normal operation, and raises alerts when there are state or transition anomalies. 

To improve the performance of these detectors against \sniper{}, we adapted  them to use only those features that are relevant to our setup. For instance, we configured \textit{Kitsune} to detect anomalies in timing, source IP, and function code combinations instead of  new communicating entity pairs.
% as in the original implementation. 
% Figure \ref{fig:detectors_overview} presents a visual overview of the alerts raised by each of these detectors during the \sniper{} attack. 
\begin{table*}[t]
\centering
% \resizebox{\textwidth}{!}{%
% \begin{tabular}{|c|c|c|c|}
% \hline
% \rowcolor[HTML]{EFEFEF} 
% Detector &
%   \begin{tabular}[c]{@{}c@{}}Type of \\ detector\end{tabular} &
%   \begin{tabular}[c]{@{}c@{}}Technique\end{tabular} &
%   \begin{tabular}[c]{@{}c@{}}Feature\end{tabular} \\
%   \hline
% \begin{tabularx}{\columnwidth}{|l|l|>{\raggedright\arraybackslash}p{3.5cm}|>{\raggedright\arraybackslash}X|}
% \renewcommand{\tabcolsep}{1.5pt}
{\renewcommand{\arraystretch}{1.2}
\begin{tabular}
{>{\raggedright\arraybackslash}p{0.15\textwidth}l
% >{\raggedright\arraybackslash}p{0.12\textwidth}
>{\raggedright\arraybackslash}p{0.33\textwidth}>{\raggedright\arraybackslash}p{0.35\textwidth}}
% {|>{\raggedright\arraybackslash}m{1.6cm}>{\raggedright\arraybackslash}m{1.4cm}>{\raggedright\arraybackslash}m{2.3cm}>{\raggedright\arraybackslash}m{2.3cm}|}

\toprule
\rowcolor[HTML]{EFEFEF}  
\textbf{Detector} & 
\textbf{Type} & 
\textbf{Technique} & 
\textbf{Feature} \\
\midrule
Inter-arrival time \cite{lin2018iat} (IaT-mean, IaT-range) & N & Statistical modeling of sample distribution (mean/range)  & Interval between consecutive occurrences of the same SCADA-\-PLC message\\
Kitsune \cite{mirsky2018kitsune} & N & Ensemble of neural networks for anomaly detection & Temporal statistics of packet sizes, count, and jitter \\ 
InvariantRules
\cite{cheng2019invariant} & P & Invariant rules for sensor/actuator values
% rule-based violations detection
& Sensor and actuator values \\
PASAD \cite{aoudi2018pasad} & P  & Singular Spectrum Analysis (SSA) & Time series of individual sensor readings \\
Histogram, MinMax, SteadyTime~\cite{wolsing2022simple} & P & Threshold-based statistical modeling trained per sensor/actuator &  Properties of process variables (e.g., value range, temporal stability) \\
\bottomrule
% \end{tabular}%
\end{tabular}
}
% }
\caption{ICS attack detectors used in our experiments. N = Network-based, P = Process log-based.}
\label{tab:detectors}
\end{table*}
\renewcommand{\tabcolsep}{6pt}

%% file: sections/evaluation.tex
\section{Evaluation}\label{sec:evaluation}
We evaluate \sniper{} on three factors that influence its success, namely profiling accuracy (\S\ref{subsec:eval-lts}), impact of attack (\S\ref{subsec:eval-attack}), and stealthiness (\S\ref{subsec:eval-detection}). We also assess its generalizability under multiple configurations of PLC periods, message transmission rates, scan cycles, and network conditions.

% To this end, we evaluate \sniper{} across the testbed configurations in Table~\ref{tab:testbed-config}, varying one or more generalization parameters at a time between consecutive configurations to understand their effect on \sniper{}'s success factors.
Table~\ref{tab:testbed-config} lists the testbed configurations used in our evaluation.
S-MD/EN-BASE and H-MD-BASE serve as the base configurations for S-SWaT and H-SWaT, respectively, using the smallest scan cycle and PLC communication periods that ensure correct process execution, as determined by industry best practices described in \S\ref{subsec:ics-state-trans}. S-EN-RNC repeats S-EN-BASE on a different day under natural network perturbations, and S-EN-PCP changes the last three PLC periods to 90 s. H-MD-LOSS1 is similar to H-MD-BASE with 1\% random packet loss added to the network. H-MD-LOSS2 also simulates the same packet loss for a different set of PLC periods.
% \aastha{Explain all configurations here.}
% \aastha{Why not use H-MD-LOSS1, H-MD-LOSS2?}
%\karthik{Maybe mention how we determined these values. }
% Across all configurations, the site-to-SCADA communication was encapsulated by an OpenVPN tunnel.

\begin{table*}[t]
\centering
\resizebox{\textwidth}{!}{%
{
\renewcommand{\arraystretch}{1.2}
\begin{tabular}{lllllll}
% {|m{1cm}|m{1cm}|m{1.5cm}|m{1.5cm}|m{1cm}|m{1cm}|m{1cm}|}
\toprule
\rowcolor[HTML]{EFEFEF}  
\textbf{Testbed} &
\textbf{Setup} &  
\textbf{Protocol} & 
\textbf{PLC Comm. periods (s)} &
\textbf{Scan cycle (s)} &
\begin{tabular}{@{}l@{}}\textbf{Message rate}\\\textbf{(msgs/s)}\end{tabular} &
\textbf{Network conditions}\\
\midrule
\multirow{4}{*}{S-SWaT} & S-MD-BASE & Unconnected Modbus TCP & 60, 60, 60, 70, 70, 70 & 60 & $\approx 40$ &No added perturbation\\
& S-EN-BASE & \cellcolor{yellow!25}Unconnected ENIP & 60, 60, 60, 70, 70, 70 & 60 & $\approx 40$ &No added perturbation\\
& S-EN-RNC & Unconnected ENIP & 60, 60, 60, 70, 70, 70 & 60 & $\approx 40$ &\cellcolor{yellow!25}\begin{tabular}{@{}l@{}}Different day,\\natural perturbations\end{tabular}\\
& S-EN-PCP & Unconnected ENIP & \cellcolor{yellow!25}60, 60, 60, 90, 90, 90 & 60 &$\approx 40$  &No added perturbation\\
\midrule
% S-EN-SCALE & \cellcolor{yellow!25}S-SWaT (12 PLCs) & Unconnected ENIP & \cellcolor{yellow!25}60, 60, 60, 60, 60, 60, 90, 90, 90, 90, 90, 90 & $\approx 60$&\cellcolor{yellow!25} $\approx 80$ &No added perturbation\\ 
% \hline
\multirow{3}{*}{H-SWaT} & H-MD-BASE & \cellcolor{yellow!25}Connected Modbus TCP & \cellcolor{yellow!25}5, 5, 10, 6, 10, 6 & \cellcolor{yellow!25}0.005 &\cellcolor{yellow!25}$\approx 65$  &No added perturbation\\
 & H-MD-LOSS1 & Connected Modbus TCP & 5, 5, 10, 6, 10, 6 & 0.005 & $\approx 65$ &\cellcolor{yellow!25}1\% packets dropped\\
 & H-MD-LOSS2 & Connected Modbus TCP & \cellcolor{yellow!25}7, 7, 10, 6, 10, 6 & 0.005 & $\approx 65$ & 1\% packets dropped\\\bottomrule
\end{tabular}
}
}
\caption{Testbed configurations for \sniper{} evaluation. Highlighted cells indicate changes from one row to the next. 
% \textcolor{red}{The width of the second-last column is unnecessarily increased by just the header. If you adjust this column's width, the table will become narrower and so will stretch better across two columns, and the font will not look so small. Tidy up these seemingly minor things to improve overall look of the tables. There is also too much duplicated text in this table. At least, swap the testbed column and setup column and use multirow to mention S-SWaT spanning the first 4 rows and H-SWaT spanning the last 3 rows only once. Remove column separators and possibly even row separators.}
}
\label{tab:testbed-config}
\end{table*}
\renewcommand{\tabcolsep}{6pt}

% We answer three questions in our evaluation.
% (i) How effectively can {\sniper} infer the SWaT testbed's superperiods and the critical messages to drop, and how well does its profiling phase generalize across diverse PLC configurations and network environments? (\S\ref{subsec:eval-lts})
% (ii) What is the impact of {\sniper}'s attack on the SWaT testbed? (\S\ref{subsec:eval-attack})
% (iii) How stealthy is {\sniper} against state-of-the-art ICS attack detection systems? (\S\ref{subsec:eval-detection}) 
% Our artifacts are available at \url{https://anonymous.4open.science/r/ics_sniper_cloud-1FEB/}

\subsection{Profiling Accuracy}
\label{subsec:eval-lts}
\sniper{}'s success in the active attack stage depends on its accuracy in the profiling stage. We assess profiling accuracy by measuring (i) accuracy in identifying superperiods, and (ii) effectiveness in identifying critical superperiods.

% To demonstrate the generalizability of \sniper{} across different PLC period configurations, ICS communication protocols, as well as to demonstrate its robustness to network conditions, we perform this evaluation across four different setups:
% \textbf{(a)} {\enipone}: testbed using ENIP/TCP with the periodicities of the 6 PLCs configured to 60 s, 60 s, 60 s, 70 s, 70 s, and 70 s. We introduce a time interval of 5s between the start times of the first 3 PLCs and the second 3 PLCs to simulate a scenario where the burst start times of all the PLCs are not perfectly aligned.
% \textbf{(b)} {\eniptwo}: the same configuration as {\enipone}, but the experiment was done on a different day to evaluate its robustness to slightly different network conditions under the same PLC configuration,
% \textbf{(c)} {\enipthree}: testbed using ENIP/TCP with the periodicities of the 6 PLCs configured to 60 s, 60 s, 60 s, 90 s, 90 s, and 90 s. This setup was used to demonstrate \sniper{}'s generalizability to different PLC periodicity combinations while using the same communication protocol. 
% \textbf{(d)} {\modbusone}: testbed using Modbus/TCP with the periodicities of the 6 PLCs configured to 60 s, 60 s, 60 s, 70 s, 70 s, and 70 s. This setup was used to demonstrate \sniper{}'s generalizability to different communication protocols.
% For all four setups, the PLC-SCADA communication was encapsulated by an OpenVPN tunnel.

\textbf{Experiment. }For each configuration, we ran our SWaT testbed for one complete operational cycle, with \sniper{} capturing the encrypted VPN traffic at VM-3 (see Figure~\ref{fig:swat-testbed}).

\textbf{Estimating superperiods.}
% \aastha{If all the results are just 100\%, is the table really informative? Might as well move it to the appendix.}\gargi{Done}
Table~\ref{tab:superperiod-result} in Appendix ~\ref{subsec:critical-superperiod} shows the actual and predicted superperiod for the seven configurations. We compute the accuracy of prediction as follows.
\begin{equation*}
    Accuracy = (1-\frac{|SP_{pred} - SP_{true}|}{SP_{true}}) \times 100\%
\end{equation*} where $SP_{true}$ is the actual superperiod and $SP_{pred}$ is the superperiod predicted by \sniper{} for a given setup. 

Across all configurations, \sniper{} achieves an accuracy of 100\% in identifying the superperiod, demonstrating generalizability. 
This holds even for H-MD-LOSS1 and H-MD-LOSS2 (different period configurations under aggressive packet loss), confirming \sniper{}'s robustness to lossy network conditions.
% \aastha{Difference between H-MD-LOSS1 and H-MD-LOSS2 is not clear from Table 2.}\gargi{Clarified and explained the configurations}
This robustness stems from \sniper{}'s use of two kinds of complementary metadata: packet size and inter-packet timing. Even when packet loss distorted the periodicities of the overlap regions ($\phi_{overlap}$) and IPT clusters ($\phi_{interval}$), each PLC still transmitted enough distinctly sized packets on schedule to preserve their periodicity, allowing \sniper{} to recover the superperiod from the $\phi_{pktlen}$ step.
% \aastha{the sentence makes contradictory claims. inter-packet timings are distorted, so how do you know the periodicity and how can you have peioridicity of packet sizes?}\gargi{Clarified}
% \karthik{Can we say why?}\gargi{Added the reason}
The hierarchical cluster boundaries are auto-derived from the level where the inter-cluster values have the largest separation, with no requirement for a manually tuned threshold, making \sniper{} portable.

% \sniper{}'s\hl{ superperiod estimation accuracy under lossy network conditions, its scalability as the number of gateway PLCs increases, and its robustness when VPN encryption renders packet sizes uniform, leaving inter-arrival timing as the only useful metadata.}  
% We present superperiod estimation accuracy under a VPN encryption scheme that renders packet sizes uniform, under lossy network conditions, and as the number of gateway PLCs increases.
% \aastha{Do not use inline bullets everywhere. Start this on a new line and end the previous sentence with a full stop instead of a colon.}

% \textit{(i) Robustness to packet losses:} We injected 1\% random packet loss on the site-to-SCADA path and evaluated \sniper{} under two additional H-SWaT period configurations: the baseline H-MD-BASE (see Table~\ref{tab:testbed-config}) and an alternative assignment of (7,7,10,6,10,6)s to PLCs 1--6, respectively.

% Post superperiod identification, \sniper{} also identified the critical transition listed in Table~\ref{tab:critical-superperiods}.
We also performed the following two supplementary evaluations of \sniper{}'s superperiod estimation.

\textit{(i) Accuracy under uniform packet sizes:} We simulate a setting where VPN encryption equalizes every packet size, thereby removing the size feature and forcing \sniper{} to rely on IPT alone. To do this, we restricted \sniper{} to only the first two timing-dependent steps of superperiod identification, i.e., $\phi_{overlap}$ and $\phi_{interval}$ (Refer to \S\ref{sec:profiling}, Step 3). We evaluated this restricted version of \sniper{} on all the S-SWaT and H-SWaT configurations listed in Table~\ref{tab:testbed-config}.
% \aastha{Is table 2 or 3 that we should be referring here? Also, did we intend to report these results for configurations like H-MD-LOSS1 and did we actually get 100\% accuracy on these? I thought we didn't and that's why we agreed that each experiment extension should be handled independently of the other extensions?}\gargi{Added the LOSS configurations back to the Tables, and mentioned that \sniper{} fails under lossy networks AND uniform packet sizes}
Across all S-SWaT configurations and H-MD-BASE, \sniper{} was able to estimate the superperiod with 100\% accuracy. For the S-SWaT configurations with low message rates, Step $\phi_{overlap}$ yielded the highest alignment score for the correct superperiod, while for the H-SWaT testbed with a high message rate, Step $\phi_{interval}$ scored the highest. However, \sniper{} could not detect the superperiod for H-MD-LOSS1 and H-MD-LOSS2, since under lossy network conditions packet length is the only feature with reliable periodicity.
% Note that a 1\% sustained packet loss is a conservative estimate. In practice, loss would be smaller and occur for short durations 
However, in real ICSes and Internet, such a sustained aggressive packet drop is rare.

% and the stable network configurations where} \sniper{} \hl{is successful are are by far the common scenarios.}
% \aastha{But say, why this is an acceptable result.}

\textit{(ii) Scalability with the number of gateway PLCs:} To evaluate superperiod detection at scale, we increased the number of gateway PLCs in the S-EN-PCP configuration from 6 to 12 by duplicating the original set of PLCs. This resulted in six PLCs with a 60 s period and six with a 90 s period, with all 12 sharing a single VPN tunnel. We found that \sniper{} estimated the superperiod (180 s) with 100\% accuracy, despite the message rate (i.e., number of messages transmitted per second between
the PLCs and SCADA over the VPN tunnel) being doubled. This robustness stems from \sniper{}'s use of three complementary metadata features: overlap region boundaries, IPTs, and packet lengths. Doubling the message rate on the VPN tunnel without lengthening the individual PLC periods obscured the overlap region boundaries, but the periodicity of individual IPT classes and distinctly sized packets from each PLC were retained, which allowed \sniper{} to recover the superperiod from the $\phi_{interval}$ and $\phi_{pktlen}$ steps.
% \aastha{Explain why you still get 100\% accuracy.}\gargi{Added}
%\karthik{Wouldn't it be quadrupled as I'd have thought there are O($n^2$) messages where $n$ is the number of gateway PLCs?}\gargi{Here, message rate denotes the number of messages transmitted per second between the six PLCs and the SCADA over the VPN tunnel, counting messages in both directions regardless of which endpoint originates them. If the six PLCs collectively account for x messages per second on average, duplicating the set to twelve PLCs yields 2x messages per second. The rate grows linearly, not quadratically, because the SCADA acts as a central hub rather than a broadcaster. Each PLC exchanges messages only with the SCADA, but the SCADA does not relay each PLC's message to every other PLC. Therefore, the O($n^2$) calculation is not applicable here.}
% \aastha{Clarify the term message rate as in your comment above, i.e., number of messages transmitted per second between the PLCs and SCADA over the VPN tunnel.}

\textbf{Identifying critical superperiods. }
% During profiling, \sniper{} identifies those superperiods as critical, whose burst structure differs from that of the preceding superperiod. In theory, burst structures of two superperiods should differ only when a state transition occurs in the underlying process. In practice, however, we fin that network jitter introduces variations in burst structures across all superperiods, especially in the high-overlap regions. To account for this, we rank the superperiods in diminishing order of their difference with the preceding superperiod. This allows \sniper{} to prioritize superperiods with the highest rank during the active attack phase. 
During profiling, \sniper{} flags a superperiod as critical if it differs from its predecessor under the two heuristics of Step 4 of \S\ref{sec:profiling}, prioritizing the one with the largest difference when several qualify. %\karthik{Is this what we mean by critical superperiod?}\gargi{Yes. Clarified.}  
Therefore, we measure \sniper{}'s effectiveness at identifying critical superperiods by the position of the first true critical superperiod on its priority list. This is an indicator of how many operational cycles \sniper{} would require to successfully disrupt a critical state transition. 
Since \sniper{} drops packets during only two superperiods per cycle, a cycle whose targeted superperiods contain no critical transition has no effect, forcing it to wait for the next cycle and the next candidate on the list. Fewer required cycles thus mean faster success and greater stealth.
% In one operational cycle, \sniper{} drops packets during only two superperiods. If the chosen superperiods 
%\karthik{there are two, not just one superperiod}\gargi{corrected} 
% in a given cycle do not contain a critical transition, the attack has no effect, and \sniper{} must wait for the next operational cycle to target the next superperiod in its priority list. Fewer required cycles thus indicate faster success and greater stealth for \sniper{}. 
In 71\% of cases, \sniper{} accurately identified the most critical superperiod as its first choice. In the remaining cases, the critical superperiod was ranked as the second most prioritized candidate by \sniper{}.
Table~\ref{tab:critical-superperiods} in Appendix \ref{subsec:critical-superperiod} shows the critical state transitions identified by \sniper{} and the number of attempts required.% for the identification. %\karthik{Do we need this table? It seems to repeat what we said here. Perhaps move it to the appendix.}\gargi{Moved the table to the Appendix}
  
%\karthik{Can we summarize the takeaway of the table in 1-2 sentences here}\gargi{Done}
% \usepackage{graphicx}
% \begin{table}[t]
% % \resizebox{\columnwidth}{!}
% \centering
% {%
% {
% % \renewcommand{\arraystretch}{1.2}
% \begin{tabular}{|l|l|l|l|l|}
% \hline
% \rowcolor[HTML]{EFEFEF} 
% \textbf{\begin{tabular}[c]{@{}c@{}}Setup\end{tabular}} &
%   \textbf{\begin{tabular}[c]{@{}c@{}}Actual SP (s)\end{tabular}} &
%   \textbf{\begin{tabular}[c]{@{}c@{}}Estimated SP (s)\end{tabular}} &
%   \textbf{\begin{tabular}[c]{@{}c@{}}Accuracy\end{tabular}}\\ \hline
% S-MD-BASE & 420 & 420 & 100\%\\ \hline
% S-EN-BASE & 420 & 420 & 100\%\\ \hline
% S-EN-RNC & 420 & 420 & 100\%\\ \hline
% S-EN-PCP & 180 & 180 & 100\%\\ \hline
% % S-EN-SCALE & 180 & 180 & 100\% & \\ \hline
% H-MD-BASE & 30 & 30 & 100\%\\ \hline
% H-MD-LOSS1 & 30 & 30 & 100\%\\ \hline
% H-MD-LOSS2 & 210 & 210 & 100\%\\ \hline
% \end{tabular}
% }%
% }
% \caption{Superperiod (SP) identification accuracy of \sniper{} under various testbed configurations. }
% \label{tab:superperiod-result}
% \end{table}

% \aastha{How are the key observations and insights connected to this shaded box here?}
\textbf{Key insights.} 
% We make the following three observations from our experiments:
\textbf{(1)} While all three event types ($\phi_{overlap}$, $\phi_{interval}$ and $\phi_{pktlen}$) used for superperiod identification (\S\ref{sec:profiling}) were effective under stable network conditions and low message rates, $\phi_{overlap}$ became unreliable under lossy network conditions, high message rates, and very short periods, which produce very short and/or inconsistent overlap regions.
% \aastha{Please mention concrete event types using symbols here.}
However, in these cases $\phi_{interval}$ and $\phi_{pktlen}$ still identified the superperiods accurately. In H-SWaT, although most payload packets were similarly sized, enough distinct sizes recurred across PLCs and states for \sniper{} to recover the superperiod.  
% \karthik{In both or just S-SWAT?} 
This confirms that our feature selection and ETA technique are robust across network conditions and PLC period configurations. \textbf{(2)} \sniper{} can successfully identify the critical superperiods where, after a critical state transition, the process remains in the new state for an extended duration, or where multiple critical transitions occur in close succession. Both of these scenarios occur in our testbeds. %\karthik{Should we say this is the case in our testbeds?}\gargi{Done}
% \textbf{(3)} Locally triggered state transitions can introduce burst pattern variations similar to those caused by remote transitions, potentially causing \sniper{} to assign higher criticality to a superperiod containing numerous local transitions. Dropping such a superperiod would be harmless, and as shown in \S\ref{subsec:eval-detection}, may not trigger any alert, giving \sniper{} another opportunity to target the next most prioritized, truly critical superperiod.
% \aastha{Why are we mentioning point 3?}

% \aastha{Use a border around insight boxes.}

\begin{tcolorbox}[colback=blue!10, colframe=black, boxrule=1pt, left=4pt, right=4pt, top=4pt, bottom=4pt, boxsep=0pt] \textbf{Summary:} \sniper{}'s process profiling technique is accurate and robust across network conditions, protocols, message rates, and PLC configurations.
\end{tcolorbox}
%\karthik{So what might happen as a result? Will it drop these packets?}\gargi{Clarified} 
% \textbf{(3)} Finally, the leakage of superperiodicity and criticality of certain sections of the OT traffic from its metadata is inherently influenced by the underlying PLC configurations and control logic, which shape the traffic burst patterns.
 
%%%%%%%%%%%%%%%%%%%%%%%%%%%%%%%%%%%%%%%%%%%%%%%%%%%%%%%%%%%%%%%%%%%%%%%%

\subsection{Attack Impact}
\label{subsec:eval-attack}
Since ICS-Sniper achieves 100\% profiling accuracy across configurations, its attack effectiveness is equally consistent. We demonstrate this on S-EN-BASE and H-MD-BASE.

\textbf{Experiment.} As mentioned in \S\ref{sec:threat-model}, \sniper{} is process-unaware, hence agnostic of the implications of the state transitions and delayed transitions on the plant. It simply drops packets whenever it predicts an upcoming critical state transition during the active attack stage. We execute one attack per operational cycle and assess its impact by inspecting whether it causes the overall process to deviate from normal behavior, resulting in reduced throughput at the end of the operational cycle, or a safety constraint violation. We study the impact of the following attacks on our testbeds: 

\textbf{Attacks on S-EN-BASE.}
% \aastha{The paragraph heading here refers to testbed, not the attack as the last sentence indicates.}
\sniper{} drops the most critical superperiod (occurring within the first 20 minutes of the operational cycle) and the second most critical superperiod (occurring within the final 30 minutes). 
We refer to these attacks as \textit{Process Delay Attack} and \textit{UF-Depletion Attack}, respectively, based on the effect that the attacks are expected to have on the underlying sub-process. \textbf{Attack on H-MD-BASE.} \sniper{} drops the most critical superperiod (occurring within the first minute of the operational cycle). This is also a variant of the process delay attack.  
% The first critical superperiod occurs within the first 20 minutes of the operational cycle, while the second critical superperiod occurs within the final 20 minutes of the operational cycle. 

\textbf{Process Delay Attack. }
In this attack, \sniper{} dropped all payload packets during the first two superperiods of the process, i.e., for 14 minutes in S-EN-BASE, and for 1 minute in H-MD-BASE. Within this time, \sniper{} dropped $\approx 4.7\%$ and $\approx 2.5\%$ of the total number of packets in S-EN-BASE and H-MD-BASE, respectively, exchanged between the site and the SCADA over the complete operational cycle. During this initial part of the operational cycle, the SCADA issues control commands to initialize PLC operations, while multiple sub-processes synchronize with one another. Packet drops within these superperiods \textit{disrupted five critical state transitions} (shown in Table \ref{tab:critical-superperiods}), stalling the operations of the sub-processes and causing a delayed process start.

\textit{Impact. }The delay caused by this attack had a cascading effect across the sub-processes, ultimately \textit{postponing the start of clean-water pumping by $\approx$ 15 minutes} in S-EN-BASE and by \textit{2 minutes} in H-MD-BASE. In H-MD-BASE, the impact cascaded further, eventually requiring 3 extra minutes of plant operation to meet its requirements.  
An average small-scale water treatment plant that pumps 100 gallons of water per minute would have a deficit of 3000 gallons and 300 gallons for S-EN-BASE and H-MD-BASE, respectively. 
% while the severity of the impact of the delay would depend on various circumstances, usually this would cause a reduction in the volume of purified water by the end of the usual operational cycle \karthik{Even an estimate would help in terms of the impact}, negatively impact downstream processes, or cause inconvenience for civilians. 

% \begin{table}[t]
% % \resizebox{\columnwidth}{!}{%
% {\renewcommand{\arraystretch}{1.2}
% \begin{tabular}{|p{2cm}|p{1cm}|p{1cm}|p{3cm}|}
% \hline
% \rowcolor[HTML]{EFEFEF} 
% \textbf{Setup} &
%   \textbf{\# superperiods} &
%   \textbf{\# attempts} &
%   \textbf{\# critical transitions} \\ \hline
% S-MD-BASE & 33 & 1 & \textbf{5} (PLC1: 1$\rightarrow$ 2, PLC2: 1$\rightarrow$ 2, PLC3: 1$\rightarrow$ 2, PLC5: 1$\rightarrow$ 3, PLC6: 1$\rightarrow$ 2)\\ \hline
% S-EN-BASE & 33 & 2 & Same as above \\ \hline
% S-EN-RNC & 32 & 2 & Same as above  \\ \hline
% S-EN-PCP & 69 & 1 & Same as above \\ \hline
% H-MD-BASE & 30 & 1 & Same as above \\ \hline
% H-MD-LOSS1 & 30 & 1 & Same as above \\ \hline
% H-MD-LOSS2 & 34 & 1 & Same as above \\ \hline
% \end{tabular}%
% }
% % }
% \caption{Critical superperiods, as identified by \sniper{}. Column 3 indicates the number of attempts required to drop the first critical superperiod.}
% \label{tab:critical-superperiods}
% \end{table}

\textbf{UF-Depletion Attack. }
In this attack, \sniper{} dropped all payload packets exchanged between the site and the SCADA during the $31^{st}$ and $32^{nd}$ superperiods of the process, i.e., for $\approx 14$ minutes. Within this time, \sniper{} dropped $\approx 4.1\%$ of the total number of packets exchanged between the site and the SCADA over the complete operational cycle. The attack stalled a critical state transition (State 13 $\rightarrow$ State 14) by disrupting synchronization between the Ultrafiltration (P3) and Backwash (P6) sub-processes. Together, they execute the ultrafiltration–backwash cycle: during backwash, P3 opens a valve to reverse flow through the UF membranes while P6 activates a pump to redirect the backwash stream to the feed tank. When P6's pump stops, P3's valve must close immediately to prevent drainage. By delaying the control message from P6 to P3 signaling the pump stoppage, the attack broke this tight coordination. 

\textit{Impact. }This attack caused the backwash valve in P3 to remain open for an \textit{additional 9 minutes and 11 seconds} after the pump in P6 stopped. This would likely result in substantial treated water loss: P6 would fail to return excess backwash water to the feed tank, and P2 would be unable to pump chemically treated water to P3 for ultrafiltration, stalling the ultrafiltration cycle and reducing overall process throughput.

\begin{tcolorbox}[colback=blue!10, colframe=black, boxrule=1pt, left=4pt, right=4pt, top=4pt, bottom=4pt, boxsep=0pt]\textbf{Summary:} All active attacks caused deviations from ideal behavior, causing degradation of plant throughput. In practice, this would incur additional operating hours or unscheduled maintenance, impose a financial burden on the plant operator, impact downstream processes, disrupt water supply to critical public services, or lower the water pressure. 
% \karthik{Seems a bit underwhelming to just talk about financial burden.}
\end{tcolorbox}
%%%%%%%%%%%%%%%%%%%%%%%%%%%%%%%%%%%%%%%%%%%%%%%%%%%%%%%%%%%%%%%%%%%%%%%%%%%%%%%%%%%%%%%%%%%%%%%%%%%%%%%%%%%%%%%%%%%%%%%%%%%%%%%%%%%%%%%%%%%%%%%%
\subsection{Stealthiness: Detecting \sniper{}}
\label{subsec:eval-detection}
\textbf{Training the detectors.}
We equipped our testbed with an ensemble of eight anomaly detectors, as described in \S\ref{sec:detectors}. They were trained with data from two complete operational cycles. The network-based detectors were trained with unencrypted traffic captured at the SCADA and the process log-based detectors were trained with sensor and actuator values collected from the PLCs, and logged by the SCADA. S-SWaT logs every 10 seconds, while H-SWaT, with a shorter scan cycle period, logs every second. We fine-tuned the parameters (e.g., window size) in the detectors such that they raise the minimum number of false alarms 
% (0\% to 21.5\% FPR, individually) 
for benign SCADA traffic. 

\textbf{Detector success.} For a detector to successfully identify an \sniper{} attack, it must detect anomalous behavior \textit{and} attribute it to the site-to-SCADA network path. This dual requirement motivates the use of both network-based and process log-based detectors. Network-based detectors alone cannot confirm process impact, as network traffic anomalies can be benign. Similarly, process log-based detectors alone cannot localize an anomaly to the network path, as sensor reading anomalies might arise due to attacks on PLCs and sensors at the site. Without both, plant operators cannot implement an effective mitigation. For instance, if only a process log-based detector raises an alert and the operator probes the site-to-SCADA link for liveness, \sniper{} would not drop probe packets unless they satisfy the ICS protocol-specific size constraints, leaving the operator unaware that the communication path is compromised.
Furthermore, an effective detector must raise an alert (i.e., detect an anomaly) as early as possible during the attack and keep raising alerts throughout the attack duration. 
% \karthik{What does remain active mean?}

We measure detector success using two metrics. \textbf{(1) Detection lag:} the time difference between the attack start time and the time of the first alert that is related to the attack event, and \textbf{(2) Alert overlap:} the time overlap between the alerts raised and the active attack window. A good detector should have a low detection lag and a high overlap.
% \karthik{Can we say what a good detector should do, have low detection lag and high overlap, is it?}

\textbf{Evaluation strategy. }To evaluate \sniper{}'s stealthiness across various generalization parameters, we measure the detector success metrics over 5 scenarios: \circled{1} Process delay attack on S-EN-BASE, \circled{2} UF-Depletion attack on S-EN-BASE, \circled{3} Process delay attack on H-MD-BASE, \circled{4} Process delay attack on H-MD-LOSS1, and \circled{5} Process delay attack on H-MD-LOSS2. 

\textbf{Results. }Table~\ref{tab:stealthiness} summarizes the performance of each detector type across these scenarios. Figures~\ref{fig:S2-networkDetectors} and \ref{fig:S2-processDetectors} in Appendix~\ref{subsec:alerts} show the efficacy of the individual and ensemble alerts of network-based and process log-based anomaly detectors, respectively, for Scenario \circled{2}. Figures~\ref{fig:S3-networkDetectors} and \ref{fig:S3-processDetectors} in Appendix~\ref{subsec:alerts} show the same for Scenario \circled{3}. In summary, while for every scenario at least one detector raised an alert at some point during the operational cycle (the ANY ensemble raised an alert), no scenario resulted in alerts from both detector types during the active attack stage (the ALL ensemble never raised an alert). For 2 out of 5 scenarios, no detector raised an alert during the active attack stage. Across all scenarios, the process underwent performance degradation before both types of detectors raised alerts. The impact was process delay ranging from 2 mins (Scenario \circled{3}) to 15 mins (Scenario \circled{1}). 

% \begin{table}[t]
% \renewcommand{\tabcolsep}{1.5pt}
% \begin{tabular}{|>{\centering\arraybackslash}p{1cm}|>{\centering\arraybackslash}p{1cm}|p{2cm}|>{\centering\arraybackslash}p{1.5cm}|>{\centering\arraybackslash}p{1.5cm}|>{\centering\arraybackslash}p{1cm}|}
% \hline
% \rowcolor[HTML]{EFEFEF}  
% \textbf{Scenario} & 
% \textbf{Active attack duration (s)} &
% \textbf{Detectors that raised alerts} & 
% \textbf{Relevance of alert to attack} & 
% \textbf{Detection Lag (s)} &
% \textbf{Alert overlap (s)}\\
% \hline
% \circled{1} & 840 & $\color{red}\bullet$IaT-mean\newline $\color{red}\bullet$IaT-range\newline $\color{red}\bullet$Kitsune\newline $\color{blue}\bullet$InvariantRules\newline $\color{blue}\bullet$PASAD\newline $\color{blue}\bullet$Histogram\newline $\color{blue}\bullet$Steady-Time & Yes (for all detectors) & 515\newline 555\newline 515\newline 514\newline 375\newline 195\newline 516& 0\newline 0\newline 0\newline 0\newline 45.5\newline 225.5\newline 0\\\hline
% \circled{2} & 840 & & & & \\\hline
% \circled{3} & 60 & & & & \\\hline
% \circled{4} & 60 & & & & \\\hline
% \circled{5} & 420 & & & & \\\hline
% \end{tabular}
% \caption{Performance of attack detectors on \sniper{}. $\color{red}\bullet$ indicates network-based detectors $\color{blue}\bullet$ indicates process log-based detectors}
% \label{tab:stealthiness}
% \end{table}
% \renewcommand{\tabcolsep}{6pt}

\begin{table}[t]
\renewcommand{\tabcolsep}{4pt}
\begin{tabular}
% {>{\centering\arraybackslash}p{1cm}>{\centering\arraybackslash}p{1cm}>{\raggedright\arraybackslash}p{2cm}>{\raggedright\arraybackslash}p{1.5cm}>{\raggedleft\arraybackslash}p{1.5cm}>{\raggedleft\arraybackslash}p{1cm}}
% {p{1.1cm}p{1.1cm}>{\raggedright\arraybackslash}p{3cm}p{0.7cm}p{1cm}p{1cm}}
{llllll}
\toprule
\rowcolor[HTML]{EFEFEF}  
\textbf{Scenario} & 
\textbf{Dur} &
\textbf{Det} & 
\textbf{Rel} & 
\textbf{Lag} &
\textbf{Overlap}\\
\midrule
\circled{1} & 840 & \begin{tabular}{@{}l@{}}$\color{red}\blacksquare$ Kitsune\\$\color{blue}\bigstar$ InvariantRules\end{tabular} &
\begin{tabular}{@{}l@{}}Yes\\Yes\end{tabular} &
\begin{tabular}{@{}l@{}}854.7\\867.7\end{tabular}& \begin{tabular}{@{}l@{}}0\\0\end{tabular}\\\midrule
\circled{2} & 840 & \begin{tabular}{@{}l@{}}$\color{red}\blacksquare$ Kitsune, IaT-mean, IaT-range\\ $\color{blue}\bigstar$ SteadyTime\end{tabular} & \begin{tabular}{@{}l@{}}Yes\\No\end{tabular} & \begin{tabular}{@{}l@{}}850.8\\--\end{tabular}& \begin{tabular}{@{}l@{}}0\\--\end{tabular}\\\midrule
\circled{3} & 60 & \begin{tabular}{@{}l@{}}$\color{red}\blacksquare$ Kitsune\\$\color{blue}\bigstar$ InvariantRules\end{tabular} &
\begin{tabular}{@{}l@{}}Yes\\Yes\end{tabular} &
\begin{tabular}{@{}l@{}}142.7\\0.4\end{tabular} & \begin{tabular}{@{}l@{}}0\\59.6\end{tabular}\\\midrule
\circled{4} & 60 & \begin{tabular}{@{}l@{}}$\color{red}\blacksquare$ Kitsune\\$\color{blue}\bigstar$ InvariantRules\end{tabular} &
\begin{tabular}{@{}l@{}}Yes\\Yes\end{tabular} &
\begin{tabular}{@{}l@{}}142.7\\0.4\end{tabular} & \begin{tabular}{@{}l@{}}0\\59.6\end{tabular}\\\midrule
\circled{5} & 420 & \begin{tabular}{@{}l@{}}$\color{red}\blacksquare$ Kitsune, IaT-mean \\$\color{blue}\bigstar$ Histogram\end{tabular} & \begin{tabular}{@{}l@{}}Yes\\Yes\end{tabular} &
\begin{tabular}{@{}l@{}}$\approx$515\\194.5\end{tabular} &
\begin{tabular}{@{}l@{}}0\\225.5\end{tabular}\\
\bottomrule
\end{tabular}
\caption{Performance of attack detectors on \sniper{}.\\$\color{red}\blacksquare$: network-based detectors, $\color{blue}\bigstar$ :  process log-based detectors. {\bf Dur}: Attack duration (s), {\bf Det}: detector with smallest detection lag and/or longest alert overlap, {\bf Rel}: relevance of alert to attack, {\bf Lag}: Detection lag (s), {\bf Overlap}: Alert overlap (s).}
\label{tab:stealthiness}
\end{table}
\renewcommand{\tabcolsep}{6pt}

% An effective detector would be able to detect attacks with high confidence and on time.  
% We evaluate the efficacy of state-of-the-art detection techniques (\S\ref{sec:detectors}) in detecting the \sniper{} attacks in terms of the following metrics: \textit{(1) Detection efficacy.} We measure the detection efficacy in terms of true positive rate (TPR) and false positive rate (FPR) (formulae in Appendix).
% \aastha{Broken reference}
% A low TPR would result in the detector missing the attack, while a high FPR would generate frequent false alarms, and increase operational costs and delays. An effective detector must have a high TPR and a low FPR. \textit{(2) Timeliness of detection. }
% A detector's timeliness is measured by the difference between the time at which the detector raised a flag and the start time of the attack. 

\textbf{Key insights. }\textbf{(1)} No network-based detector raised an alert until the attack had concluded. This is due to their fundamental design characteristic: their anomaly detection logic is event-driven, triggering only upon packet arrival, leaving even prolonged silent periods on the communication link undetected. This design suits volumetric DoS attacks, where it enables real-time response under high packet rates, but leaves the detectors blind to blackhole attacks like \sniper{}.
These findings highlight the need for a new detection approach for targeted blackhole attacks on ICS networks. Detecting such attacks would instead require periodically polling the communication link, synchronized to PLC periodicities and network conditions, which is a non-trivial research challenge.

\textbf{(2)} Unlike H-SWaT, no process log-based detector raised alerts in S-SWaT. This is due to S-SWaT's longer scan cycle and 10$\times$ slower logging rate,  which reduce observability of process value drifts and prevent detectors from triggering. %capturing the attack's impact.

\textbf{(3)} Detector performance is best under stable network conditions ($<$ 1\% packet drops) and short PLC communication periods (Scenarios \circled{3} and \circled{4}). This is because longer periods in a lossy network slow the message rate, making deviations harder to detect. 

\begin{tcolorbox}[colback=blue!10, colframe=black, boxrule=1pt, left=4pt, right=4pt, top=4pt, bottom=4pt, boxsep=0pt]\textbf{Summary:} No detector could effectively detect \sniper{} during its active attack window across all generalization parameters.\end{tcolorbox}

%% file: sections/limitations_countermeasures.tex
\section{Discussion}\label{sec:countermeasures}
\textbf{Impact of sporadic manual commands and out-of-routine state changes.} These factors do not mislead \sniper{}, because its profiling stage repeats the traffic-capture step over multiple operational cycles until a consistent per-cycle packet count recurs, then selects a representative sequence. Therefore, cycles perturbed by such events are discarded and never drive prediction. These events might raise \sniper{}'s monitoring cost, but a nation-state actor %would plausibly 
may have the resources to stealthily 
monitor the ICS for a long time before attacking.

\textbf{Limitation of \sniper{}.}
\sniper{}'s efficacy relies on the difference in IPTs between a batch-processing ICS's operational and idle phases. Continuous-processing ICSes lack distinct operational and idle phases, which may render ICS-Sniper ineffective against them. However, many safety-critical ICSes are batch-processing systems~\cite{batch-water, batch-processing}, against which \sniper{} remains effective. 
% \aastha{Broken reference}

% Second, {\sniper} has currently been tested on 6 PLCs. In practical scenarios, ICSes may have 10-100 PLCs in each process that share a VPN tunnel. Identifying critical messages in such a setup would require enhancing {\sniper}'s pattern mining algorithm.  
% \karthik{how exactly?}\gargi{Removing this, as it is not true according to our latest findings.}
% In practical scenarios, ICSes may have multiple PLCs in each sub-process that communicate in parallell with other sub-processes. Identifying critical messages in such a setup would require enhancing {\sniper}'s pattern mining algorithm. However, this extension should be straightforward.

\textbf{Countermeasures.}
%\hltk{We should say explicitly that the attack is easy to prevent, but that doesn't take away the contribution of the paper as we're the first to identify it. This is what reviewer 2 asked.}
% \rev{}
% \textbf{Improving detectors.} -- Correlating repeated network outages with the sensitivity of the ICS communications affected might provide valuable insights regarding the nature of the network outage. Such correlation techniques would aid the development of stronger intrusion detection systems (IDS). However, developing such an IDS would require a lot of data points, which means the ICS has to withstand a significant number of network outages before its IDS can be strengthened. 
% A potential workaround for this would be to use a honeypot that mimics the original ICS communications. A honeypot can be used to check the sanity of the network path prior to sending the original sensitive ICS messages. If the honeypot experiences network fluctuations, the timing of the fluctuations can be correlated with the sensitivity of the honeypot messages for detecting the presence of an adversary on the path. It would still be a challenge to ensure that the honeypot messages follow the exact same path as the ICS.
% \am{Since we showed the ineffectiveness of detection techniques, I wonder if we should propose another IDS solution.}\hltg{\textbf{GM: I think we can either omit it or suggest as a future work that ICSes need a more sophisticated IDS that can identify targeted packet drops from outside the perimeter. We can stop at that itself.}}
\sniper{} would not succeed if no on-path network device or service had an exploitable vulnerability. However, securing every network device outside the ICS perimeter is highly challenging, as it requires collaboration with ISPs and a deep understanding of the impact of packet drops at different points during the operational cycle. 
Therefore, we consider other countermeasures. 
%We therefore focus on countermeasures that an ICS can adopt to protect its traffic even under such a compromise.

\textit{Traffic shaping.}
{\sniper} relies on periodic burst patterns in PLC traffic to predict state transitions. Therefore, a principled approach to mitigating the attacks would require {\em shaping} the traffic to obfuscate the periodic burst patterns. Traffic shaping can be implemented at either the application or network layer; however, PLC periodicities and burst durations are typically tuned for optimal performance. Designing a shaping strategy that ensures security without sacrificing performance remains an open research challenge.

\textit{Redundant network paths.}
% To enhance resilience against \sniper{},
ICSes could route their OT traffic over multiple Internet paths, e.g., using source routing~\cite{barrera2017scion}.

%% file: sections/relwork.tex
\section{Related Work}\label{sec:rel-work}

\textbf{Encrypted Traffic Analysis-based Attacks. }
Encrypted Traffic Analysis (ETA) has been used for inferring users' website visits~\cite{wang2014effective} and IoT device activity~\cite{papadogiannaki2021survey,msadek2019iot}, where an adversary first profiles the traffic by interacting with the target as a legitimate user. However, ICS environments lack a public-facing interface, making such interactions, and thus detailed behavioral profiling, infeasible for an external adversary. Existing work has only coarsely distinguished encrypted ICS control traffic from general Internet traffic~\cite{barbieri2021assessing}, without conducting fine-grained analysis of ICS message criticality.

\textbf{Existing ICS DoS Attacks.}
DoS attacks on ICSes can be classified based on the adversary's vantage point. (i) \textit{Data modification attacks} involve infiltrating ICS networks to modify or drop time-sensitive sensor readings, disrupting critical operations~\cite{krotofil2014cps, ndonda2022exploiting}. (ii) \textit{Traffic flow modification attacks} are carried out by adversaries who cannot infiltrate the ICS network and thus cannot observe packet contents to assess their importance. These adversaries may use two main tactics: (i) \textit{Volumetric DoS attacks} that gradually overload a service with illegitimate requests, preventing it from processing legitimate ones~\cite{gasser2017amplification}, and (ii) \textit{Blackhole attacks} that disable a target service by dropping all traffic sent to it~\cite{mahmood2007survey, pham2015detecting, airehrour2016securing, abosata2021internet}. 

\textbf{DoS detection techniques. }Apart from the ICS DoS detection techniques in \S\ref{sec:detectors}, there are blackhole attack detectors and conventional cloud-based DoS detectors. Blackhole attack detectors~\cite{mahmood2007survey,airehrour2016securing,abosata2021internet} focus on anomalies in route discovery, but struggle with attacks that only drop a few packets. Conventional cloud defenses target ARP poisoning, payload tampering, and high-volume floods through implementing static ARP, endpoint security, and high-volume-flow detection~\cite{aws-shield,reece2024defending,fayaz2015bohatei}. ICS-Sniper introduces no malicious ARP, alters no payload, and generates no extra traffic, and only selectively drops a few packets. Hence, while these defenses are effective in securing the web services that dominate cloud traffic, they are not designed to detect ICS-Sniper.

% There are two broad categories of ICS DoS detection techniques. (i) \textit{Network traffic-based anomaly detection:} These techniques~\cite{lin2018iat, mirsky2018kitsune} identify traffic-flow modification attacks by comparing real-time traffic with models built under normal conditions. %The challenge is setting thresholds to differentiate between malicious deviations and normal fluctuations. %In MANETs, 
% blackhole attack detectors~\cite{mahmood2007survey,airehrour2016securing,abosata2021internet} focus on anomalies in route discovery, but struggle with attacks that only drop a few packets.
% (ii) \textit{Process log-based anomaly detection:} These techniques~\cite{cheng2019invariant, aoudi2018pasad, kim2020seq2seq, wolsing2022simple} model ICS behavior by correlating sensor readings and process variables, then compare real-time data with the model to detect anomalies. The challenge in these techniques is in setting thresholds— high thresholds might miss attacks or delay detection, while low thresholds could cause false alarms due to natural fluctuations.
  
% {\sniper} is a targeted blackhole DoS attack that can impact an ICS by dropping a very low volume of ICS traffic and thus evades detection by state-of-the-art detectors.

%% file: sections/conclusion.tex
\section{Conclusion}\label{sec:conclusion}
We present \emph{{\sniper}, the first targeted blackhole DoS attack technique for modern batch-processing Industrial Control Systems (ICS) that host the SCADA on the cloud.}
% We present \sniper{}, a targeted blackhole attack on geo-distributed batch-processing ICSes.
\sniper{} leverages the periodic and bursty nature of ICS communications and the correlation between packet metadata and ICS states to infer underlying state transitions and critical messages. It can attack an ICS from a compromised ISP device outside the ICS perimeter, causing substantial damage while remaining undetected.
Our case study on a water treatment ICS testbed demonstrates that \sniper{} can potentially cause loss of productivity of the ICS. As cloud-hosted SCADA becomes a necessity, our findings underscore the need for stronger protections in modern ICSes beyond the NIST recommendations. To mitigate \sniper{}, we recommend strong countermeasures based on traffic shaping and redundancy in routing.

Future work will focus on expanding \sniper{} to a larger class of ICSes, including continuous processing ones. We will also explore mitigation strategies against the attacks.

% \aastha{This sentence should come later.}

%% file: appendix.tex
\subsection{Sample control algorithm}\label{subsec:control-algo}
\begin{figure*}[t!]
\centering
\begin{subfigure}{.5\textwidth}
\begin{lstlisting}[style=ST, xleftmargin=15pt]
PROGRAM P1_Control
//Initializing parameters
VAR
    // Overall plant status
    plantOn: BOOL := False; 
    // Sub-process state
    p1State: INT := 1;
    // device configurations
    pmp1Status: BOOL := False;
    vlv1Status: BOOL := False;
    vlv2Status: BOOL := False;
    //sensor values
    lvl1Val: INT := Low;
END_VAR
// Main block
IF p1State==1 THEN
    // Obtain remote input from SCADA
    GETVAL(plantOn) 
    GETVAL(vlv2Status)
    // Control logic
    IF plantOn THEN
        IF vlv2Status==True AND lvl1Val>Low THEN
            pmp1Status := True;
        END_IF    
        IF lvl1Val<High THEN
            vlv1Status := True; 
        END_IF    
        IF lvl1Val==High THEN
            vlv1Status := False;
        END_IF
        IF vlv2Status==False THEN
            // State transition
            p1State := 2;
        END_IF    
    END_IF 
    // Send data updates to SCADA
    SETVAL(lvl1Val) 
    SETVAL(pmp1Status)
    SETVAL(vlv1Status)    
END_IF
IF p1State==2 THEN
    // Obtain remote inputs from SCADA
    GETVAL(plantOn)
    // Control logic
    pmp1Status := False
    IF !plantOn OR lvl1Val==High THEN
        vlv1Status := False;
        // State transition
        p1State := 1;
    ELSE
        vlv1Status := True;
    END_IF
    // Send data updates to SCADA
    SETVAL(lvl1Val) 
    SETVAL(pmp1Status)
    SETVAL(vlv1Status)    
END_IF 
\end{lstlisting}
\end{subfigure}%
\hfill
\begin{subfigure}{.5\textwidth}
\begin{lstlisting}[style=ST, xleftmargin=18pt]
PROGRAM P2_Control
//Initializing devices
VAR
    // Overall plant status
    plantOn: BOOL := False; 
    // Sub-process states
    p2State: INT := 1;
    // device configurations
    pmp2Status: BOOL := False;
    vlv2Status: BOOL := False;
    //sensor values
    lvl2Val: INT := Low;
    flw2Val: INT := 0;
END_VAR
// Main block
IF p2State==1 THEN
    // Obtain remote input from SCADA
    GETVAL(plantOn) 
    // Control logic
    IF plantOn THEN
        vlv2Status := True
        IF lvl2Val>Low AND lvl2Val<High AND flw2Val<W THEN   
            pmp2Status := True; 
        END_IF    
        IF lvl2Val==High THEN
            // State transition
            p2State := 2;
        END_IF
    ELSE
        pmp2Status := False;
        vlv2Status := False;
    END_IF    
    // Send data updates to SCADA
    SETVAL(lvl2Val) 
    SETVAL(pmp2Status)    
    SETVAL(vlv2Status)
    SETVAL(flw2Val) 
END_IF
IF p2State==2 THEN
    // Obtain remote inputs from SCADA
    GETVAL(plantOn)
    // Control logic
    vlv2Status := False;
    IF flw2Val<W THEN
        pmp2Status := True;
    END_IF    
    IF flw2Val==W OR !plantOn THEN
        pmp2Status := False;
        vlv2Status := False;
        // State transition
        p2State := 1;
    END_IF
    // Send data updates to SCADA
    SETVAL(lvl2Val) 
    SETVAL(pmp2Status)
    SETVAL(vlv2Status)
    SETVAL(flw2Val) 
END_IF
\end{lstlisting}
\end{subfigure}%
\caption{Control algorithms for a sample ICS described in Section~\ref{sec:prelims}.}
\label{fig:samplecontrolalg}
\end{figure*}

\newpage
\subsection{Corner case analysis of the profiling stage of \sniper{} }\label{subsec:corner-case}

\sniper{} is likely to mispredict state transitions or fail to identify the superperiod in the following cases. 
% \karthik{Maybe move this to the appendix/}
\begin{enumerate}[wide, labelwidth=5pt, labelindent=8pt]
    \item \sniper{} infers state transitions from changes in traffic burst patterns, but it cannot distinguish whether a transition was triggered by a local sensor input or by remote sub-process input. If all the transitions in the identified superperiods are local, then dropping packets will not affect the process, even though the transitions are detected. In such cases, the attack would have a limited impact. 
    % \item \sniper{} prioritizes stable regions (those with minimal or no overlapping bursts) for identifying the dominant period. Theoretically, such regions might not exist if the periodicities and burst phases of all PLCs align so that multiple bursts always overlap. However, in practice, this scenario is highly unlikely, since individual PLC burst durations are always shorter than their respective periods, resulting in a short period of network inactivity every period for all PLCs. Therefore, at least some non-overlapping (stable) regions must almost always occur. 
    \item Within a superperiod, if multiple sub-processes undergo state transitions such that the aggregate traffic metadata before and after the transitions remains the same, \sniper{} would not detect the transitions. While possible, such scenarios would be unusual and would not likely be frequent in practice.
    \item If all sub-processes undergo very frequent state transitions, i.e., a large number of per-PLC state transitions inside each superperiod and in every superperiod over the length of the operational cycle, it would be challenging for \sniper{} to identify a dominant period.
    \item We know that the superperiod L is the LCM of the periods of the constituent PLCs. Theoretically, L can grow very large when PLC periods include a large number of mutually prime factors. If L exceeds one-third of the length of the operational cycle, \sniper{} cannot observe repetition of a full superperiod, and therefore cannot compute it. However, in practice, PLC communication cycles are in the order of seconds while operational cycles span several hours, so this scenario is highly unlikely.  
\end{enumerate}

%%%%%%%%%%%%%%%%%%%%%%%%%%%%%%%%%%%%%%%%%%%%%%%%%%%%%%%%%%%%%%%%%%%%%%%%%%%%%%%%%%%%%%%%%%%%%%%%%%%%%%%%%%

\subsection{Comparison with volumetric blackhole attacks}\label{sec:comparison}
Individual PLC periods are on the order of seconds, far shorter than the operational cycle, which may span hours to a day. The superperiods are therefore also on the order of seconds to a few minutes. Dropping packets in a targeted way within just two small windows thus makes \sniper{} stealthier than a conventional volumetric blackhole attack.
% Since the periods of individual PLCs are in the order of seconds, which is much smaller compared to the overall operational cycle time, which might be in the order of several hours to a day. Consequently, the superperiods are also in the order of seconds to a few minutes, much smaller as compared to the operational cycle length.
% Therefore, dropping packets~in~a targeted manner within two small time windows makes \sniper{} stealthier than conventional volumetric blackhole attacks.
We compare the stealth of \sniper{} with a non-targeted volumetric blackhole attack in terms of the number of packets they drop.

\textbf{Number of packets dropped by \sniper{}. }On detecting a burst-pattern change in superperiod x, \sniper{} drops all payload packets in superperiods (x-1) and x. The total number of dropped packets is 
\begin{equation}
    Drops_{\sniper{}} = f\% \text{ of }\sum_{i=1}^{n}\sum_{j=\frac{L(x-2)}{P_i}+1}^{\frac{Lx}{P_i}}pkt_{ij}
\end{equation}
where n is the number of PLCs, L is the superperiod, $P_i$ is the period of s(i), and $pkt_{ij}$ is the number of packets of s(i) in its $j^{th}$ period, f indicates the percentage of application payload packets within the time window. f=20 for unconnected protocol variants, and f=100 for connected protocol variants. 

\textbf{Worst-case comparison with a naive adversary. }A naive adversary not equipped with \sniper{} knows neither L nor the payload packet sizes, and therefore begins dropping packets from the start of the operational cycle, and continues until it drops the first critical packet. If the first critical packet occurs in superperiod x, the \textit{minimum} number of packets it drops (the adversary can drop any number of packets even after that) is
\begin{equation}
    Drops_{naive} = \sum_{i=1}^{n}\sum_{j=1}^{\frac{L(x-1)}{P_i}}pkt_{ij} + c, 
\end{equation}
where c is the number of packets in the $x^{th}$ superperiod until the critical packet. 
Comparing equations (1) and (2), we see that, for unconnected protocol variants \sniper{} drops at least $80\%$ fewer packets than a naive volumetric blackhole attacker. In the connected variant, the two would drop the same number of packets only when the second superperiod is critical, and the volumetric attack stops after the second superperiod.  
The stealth of \sniper{} becomes more significant as the value of x increases, i.e., for state changes at later stages in the operational cycle, because prolonged indiscriminate packet drops by a naive adversary could be detected with network polling packets used for liveness checks. \sniper{} would not drop these polling packets.

\textbf{Best-case comparison with a naive adversary. }In the best-case scenario, the naive adversary picks a random start time and duration and, by chance, drops only the critical packets. Without knowledge of L, it would have infinitely many choices over the entire round of operation. On the other hand, if an \sniper{}-equipped adversary wants to pick a random start time and duration for packet drops, they would do so only within the $(x-1)^{th}$ and $x^{th}$ superperiods. Therefore, the naive adversary has a much lower success probability than that of \sniper{} due to \sniper{}'s much smaller search window. 

%%%%%%%%%%%%%%%%%%%%%%%%%%%%%%%%%%%%%%%%%%%%%%%%%%%%%%%%%%%%%%%%%%%%%%%%%%%%%%%%%%%%%%%%%%%%%%%%%%%%%%%%%%
\subsection{Supplementary information about the SWaT testbed}\label{subsec:swat-supplementary}
\begin{figure}[h]
    \begin{center}
    \includegraphics[scale=0.5]{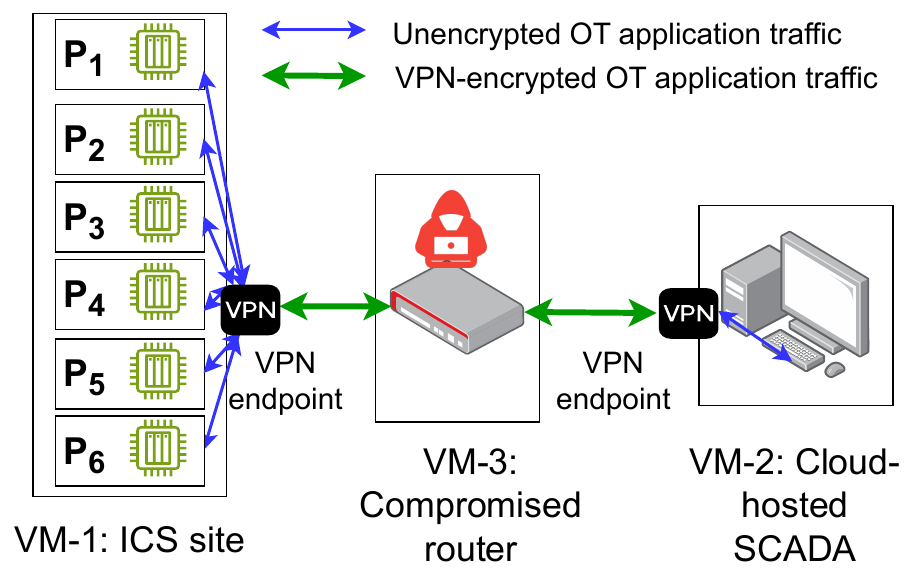}
    % \captionsetup{justification=centering}
    \caption{A schematic diagram of our SWaT testbed}
    \label{fig:swat-testbed}
    \end{center}
\end{figure}  

\begin{figure}[h!]
    \begin{center}
    \includegraphics[scale=0.07]{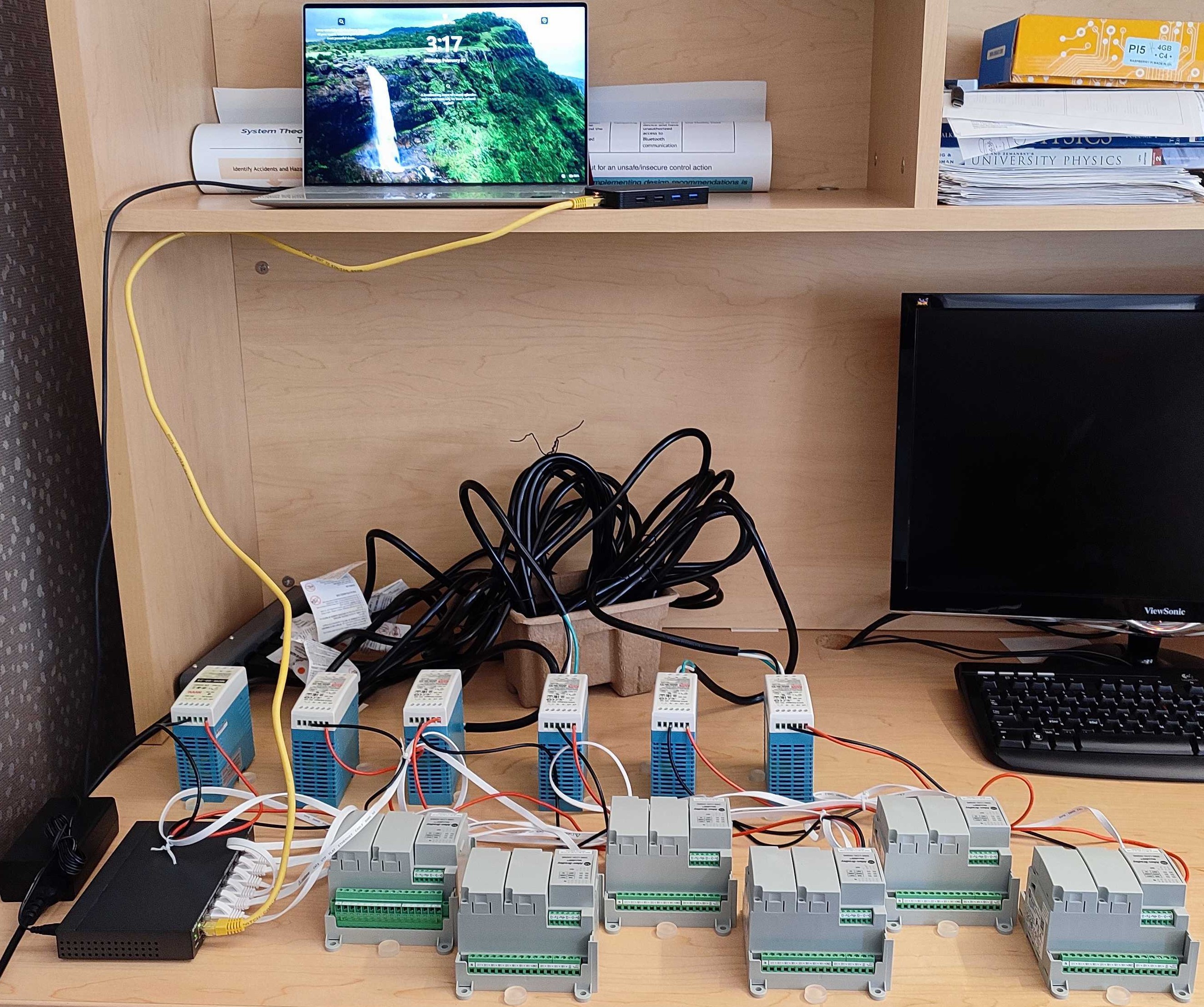}
    % \captionsetup{justification=centering}
    \caption{The site network of our H-SWaT testbed. The six PLCs represent the six sub-processes. The laptop acts as a gateway router and also runs the physical process simulation. The compromised router with \sniper{} code and the SCADA are deployed on two different EC2 VMs.}
    \label{fig:hswat}
    \end{center}
\end{figure}
% \newpage

\begin{table}[h]
% \resizebox{\columnwidth}{!}{%
\begin{tabular}{lllll}
\toprule
\rowcolor[HTML]{EFEFEF} 
\textbf{Sub-process} &
\textbf{\begin{tabular}[c]{@{}c@{}}\# states\end{tabular}} &
\textbf{\begin{tabular}[c]{@{}c@{}}R\end{tabular}} &
\textbf{\begin{tabular}[c]{@{}c@{}}\# critical msgs\end{tabular}} &
\textbf{\begin{tabular}[c]{@{}c@{}}\# non-critical msgs\end{tabular}}
\\
\midrule
P1 & 3  & 2  & 3 & 41  \\
P2 & 2  & 2  & 2 & 100 \\
P3 & 20 & 24 & 4 & 60  \\
P4 & 6  & 1  & 1 & 80  \\
P5 & 21 & 21 & 3 & 60  \\
P6 & 2  & 2  & 3 & 69  \\
\bottomrule
\end{tabular}%
% }
\caption{Information about the state transition model of the SWaT testbed. {\bf R:} \# of remotely triggered state transitions.}
\label{tab:testbed}
\end{table}

%%%%%%%%%%%%%%%%%%%%%%%%%%%%%%%%%%%%%%%%%%%%%%%%%%%%%%%%%%%%%%%%%%%%%%%%%%%%%%%%%%%%%%%%%%%%%%%%%%%%%%%%%%
\newpage
\subsection{Superperiod identification by \sniper{}}\label{subsec:critical-superperiod}
\begin{table}[h]
\centering
% \resizebox{\columnwidth}{!}
% {%
{
\begin{tabular}{llll}
\toprule
\rowcolor[HTML]{EFEFEF} 
\textbf{\begin{tabular}[c]{@{}c@{}}Setup\end{tabular}} &
  \textbf{\begin{tabular}[c]{@{}c@{}}Actual SP (s)\end{tabular}} &
  \textbf{\begin{tabular}[c]{@{}c@{}}Estimated SP (s)\end{tabular}} &
  \textbf{\begin{tabular}[c]{@{}c@{}}Accuracy\end{tabular}}\\ \midrule
S-MD-BASE & 420 & 420 & 100\%\\
S-EN-BASE & 420 & 420 & 100\%\\
S-EN-RNC & 420 & 420 & 100\%\\
S-EN-PCP & 180 & 180 & 100\%\\
% S-EN-SCALE & 180 & 180 & 100\% & \\ \hline
H-MD-BASE & 30 & 30 & 100\%\\
H-MD-LOSS1 & 30 & 30 & 100\%\\
H-MD-LOSS2 & 210 & 210 & 100\%\\
\bottomrule
\end{tabular}
}%
% }
\caption{Superperiod (SP) identification accuracy of \sniper{} under various testbed configurations. }
\label{tab:superperiod-result}
\end{table}

\begin{table}[h]
% \resizebox{\columnwidth}{!}{%
{
\begin{tabular}{p{2.2cm}p{1cm}>{\raggedright\arraybackslash}p{1.3cm}>{\raggedright\arraybackslash}p{2.5cm}}
\toprule
\rowcolor[HTML]{EFEFEF} 
\textbf{Setup} &
  \textbf{\# superperiods} &
  \textbf{\# attempts} &
  \textbf{\# critical transitions} \\ \midrule
S-MD-BASE & 33 & 1 & \textbf{5} (PLC1: 1$\rightarrow$ 2, PLC2: 1$\rightarrow$ 2, PLC3: 1$\rightarrow$ 2, PLC5: 1$\rightarrow$ 3, PLC6: 1$\rightarrow$ 2)\\
S-EN-BASE & 33 & 2 & Same as above \\
S-EN-RNC & 32 & 2 & Same as above  \\
S-EN-PCP & 69 & 1 & Same as above \\
H-MD-BASE & 239 & 1 & Same as above \\
H-MD-LOSS1 & 239 & 1 & Same as above \\
H-MD-LOSS2 & 34 & 1 & Same as above \\ 
\bottomrule
\end{tabular}%
}
% }
\caption{Critical superperiods, as identified by \sniper{}. Column 3 indicates the number of attempts required to drop the first critical superperiod.}
\label{tab:critical-superperiods}
\end{table}

%%%%%%%%%%%%%%%%%%%%%%%%%%%%%%%%%%%%%%%%%%%%%%%%%%%%%%%%%%%%%%%%%%%%%%%%%%%%%%%%%%%%%%%%%%%%%%%%%%%%%%%%%%

\subsection{Alerts generated by anomaly detectors}\label{subsec:alerts}
Throughout this subsection, red markers denote alerts outside the active attack window, whereas black markers denote alerts overlapping with the active attack window.
\begin{figure}[h]
    \centering
    % \begin{center}
    \includegraphics[scale=0.7]{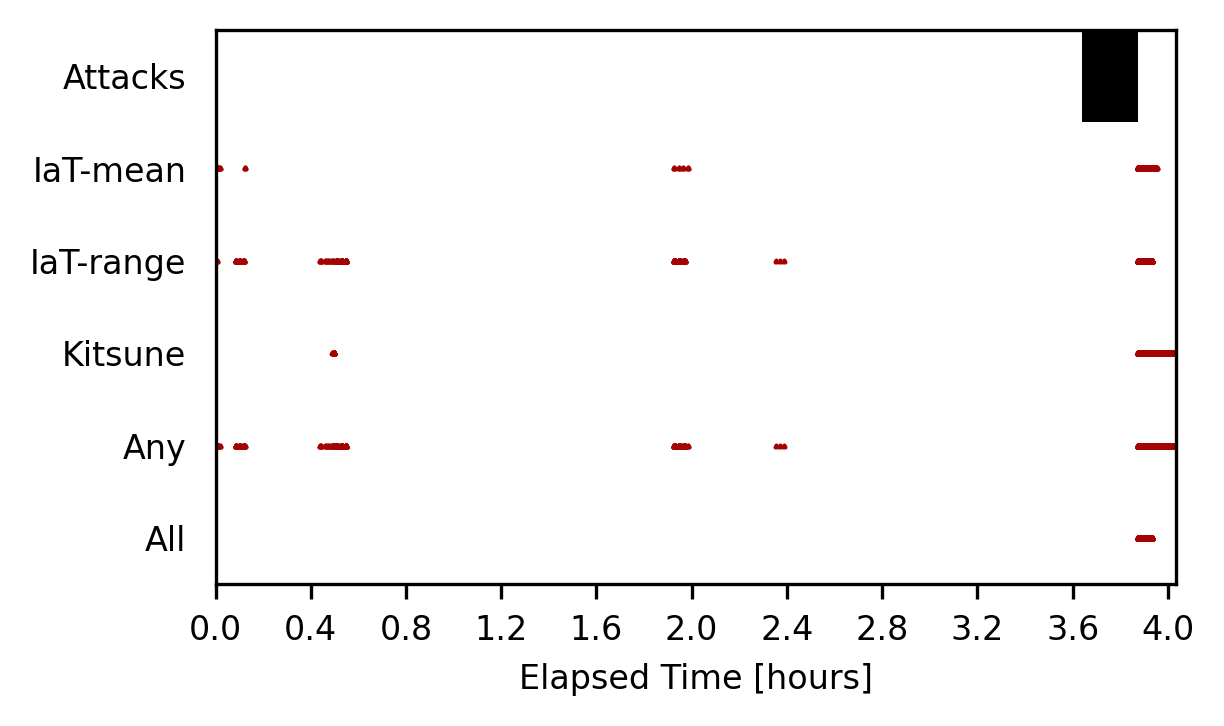}
    % \captionsetup{justification=centering}
    \caption{Alerts generated by the network-based anomaly detectors over time for Scenario \protect\circled{2}. 
    % Red: Alerts outside of the active attack window, Black: Alerts overlapping with the active attack window.
    }
    \label{fig:S2-networkDetectors}
    % \end{center}
\end{figure}

\begin{figure}[h]
    \centering
    % \begin{center}
    \includegraphics[scale=0.7]{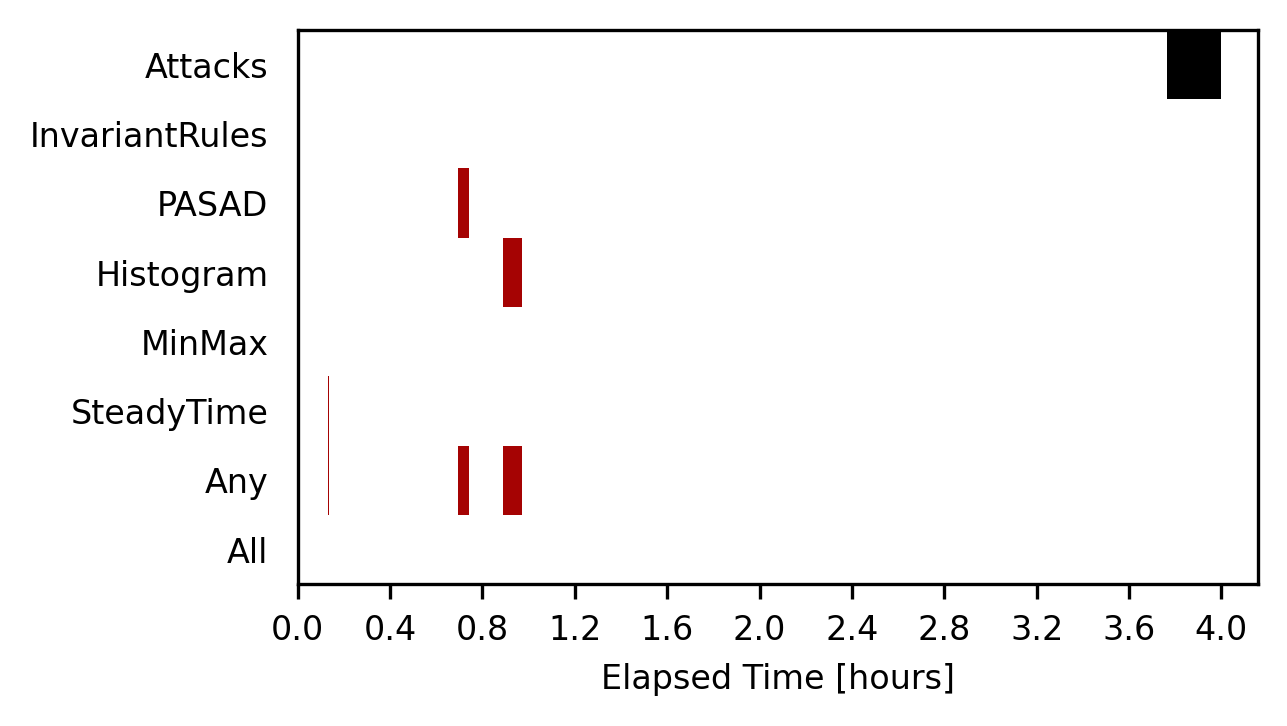}
    % \captionsetup{justification=centering}
    \caption{Alerts generated by the process log-based anomaly detectors over time for Scenario \protect\circled{2}. 
    % Red: Alerts outside of the active attack window, Black: Alerts overlapping with the active attack window.
    }
    \label{fig:S2-processDetectors}
    % \end{center}
\end{figure}

\begin{figure}[h]
    \centering
    % \begin{center}
    \includegraphics[scale=0.7]{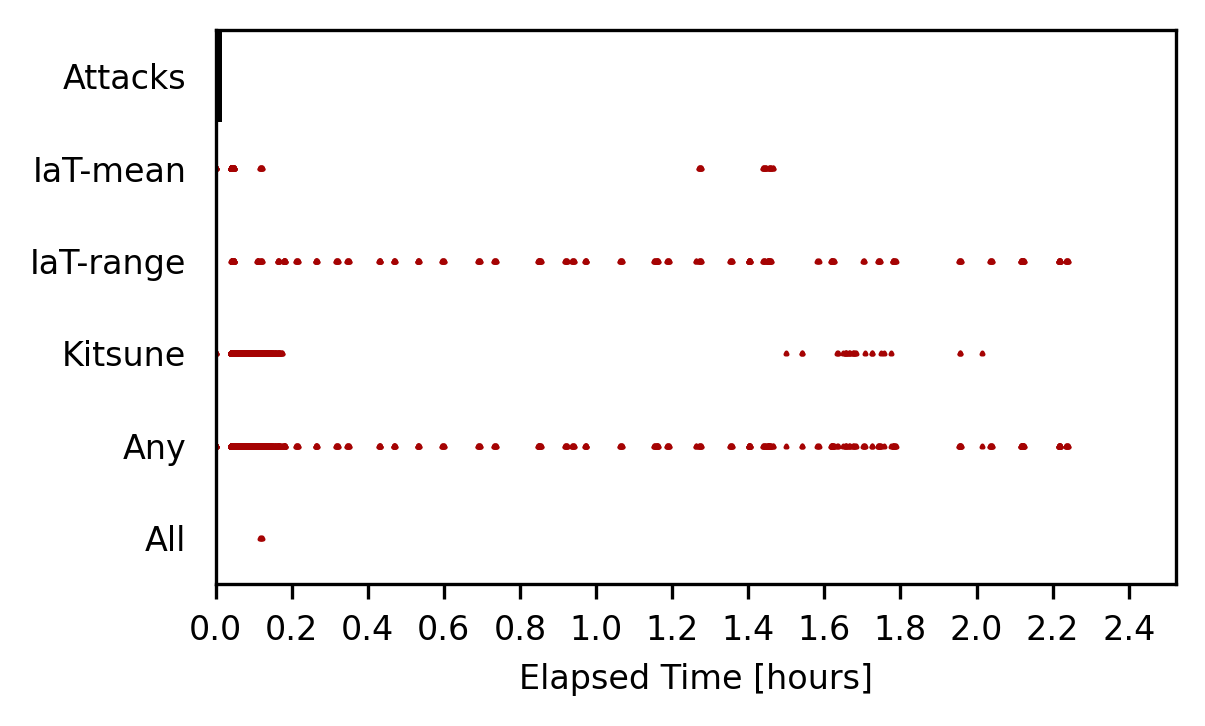}
    % \captionsetup{justification=centering}
    \caption{Alerts generated by the network-based anomaly detectors over time for Scenario \protect\circled{3}. 
    % Red: Alerts outside of the active attack window, Black: Alerts overlapping with the active attack window.
    }
    \label{fig:S3-networkDetectors}
    % \end{center}
\end{figure}

\begin{figure}[h]
    \centering
    % \begin{center}
    \includegraphics[scale=0.7]{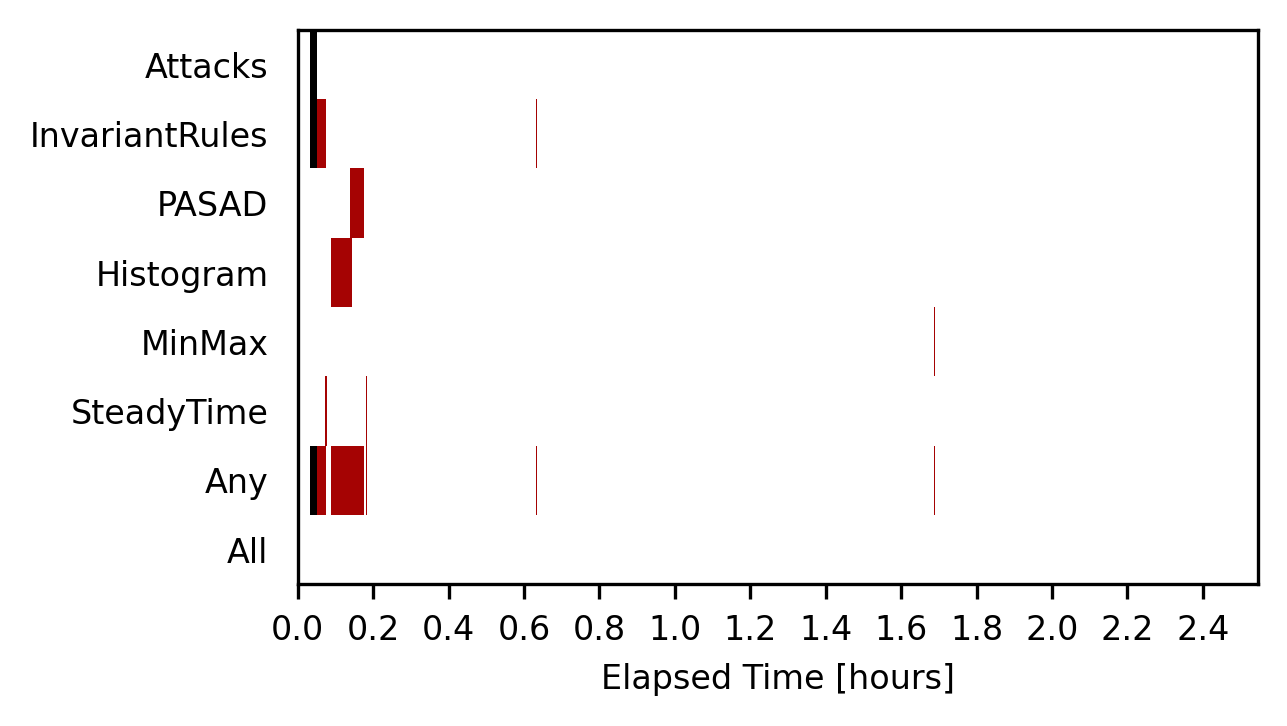}
    % \captionsetup{justification=centering}
    \caption{Alerts generated by the process log-based anomaly detectors over time for Scenario \protect\circled{3}. 
    % Red: Alerts outside of the active attack window, Black: Alerts overlapping with the active attack window.
    }
    \label{fig:S3-processDetectors}
    % \end{center}
\end{figure}

%% file: artifact-appendix.tex
\appendices
\section{Artifact Appendix}

\subsection{Description \& Requirements}
This artifact accompanies the paper \textit{ICS-Sniper: A Targeted Blackhole Attack on Encrypted ICS Traffic}. The package contains \textit{S-SWaT}, a software-simulated six-stage water treatment plant (Python control logic and process models) across three cloud VMs (plant site, cloud SCADA, on-path router), with configurable PLC periods, protocols (Modbus/TCP, ENIP/TCP), and network conditions; \textit{H-SWaT} deployment instructions and structured-text control logic for Allen-Bradley Micro820 PLCs with the SCADA; \textit{encrypted VPN traces} from both testbeds for all configurations; the \textit{ICS-Sniper implementation} (process profiling and active attack modules); and \textit{IPAL}, an eight-detector ensemble adapted to the ICS scenario. 

\subsubsection{How to access}
% The artifact is organized into 5 public GitHub repositories and a Google Drive link containing the captured traces. Each repository has its own README file, describing how to set up the testbeds, execute the code, and verify the results.
The artifact is archived in a single Zenodo record (DOI: \href{https://zenodo.org/records/23737012}{10.5281/zenodo.23737011}), which bundles the five components listed in Table~\ref{tab:artifacts}, each archived as a .zip version of its public GitHub repository at tag v1.0-artifact. The Zenodo record contains a top-level README, a LICENSE, and checksum/provenance files (SHA256SUMS.txt and
COMMITS.txt). Each repository also has its own README describing how to set up the testbeds, execute the code, and verify the results.

% \begin{itemize}
%     \item \texttt{Modbus-C-hardware-testbed}: This repository contains the implementation of the H-SWaT testbed using the Connected Modbus protocol. GitHub URL: \href{https://github.com/ICS-Sniper/Modbus-C-hardware-testbed.git}{https://github.com/ICS-Sniper/Modbus-C-hardware-testbed.git}  
%     \item \texttt{Modbus-U-software-testbed}:  This repository contains the implementation of the S-SWaT testbed using the Unconnected Modbus protocol. GitHub URL: \href{https://github.com/ICS-Sniper/Modbus-U-software-testbed.git}{https://github.com/ICS-Sniper/Modbus-U-software-testbed.git}
%     \item \texttt{ENIP-U-software-testbed}: This repository contains the implementation of the S-SWaT testbed using the Unconnected ENIP protocol. GitHub URL: \href{https://github.com/ICS-Sniper/ENIP-U-software-testbed.git}{https://github.com/ICS-Sniper/ENIP-U-software-testbed.git}
%     \item \texttt{ICS-SNIPER-attack}: This repository contains the implementation of ICS-Sniper. GitHub URL: \href{https://github.com/ICS-Sniper/ICS-SNIPER-attack.git}{https://github.com/ICS-Sniper/ICS-SNIPER-attack.git}
%     \item \texttt{SOTA-detectors-NDSS}: This repository contains the implementation of IPAL detectors, adapted to ICS traffic, with preprocessed datasets, and code for evaluation and plotting. GitHub URL: \href{https://github.com/ICS-Sniper/SOTA-detectors-NDSS.git}{https://github.com/ICS-Sniper/SOTA-detectors-NDSS.git}
%     \item \texttt{Dataset}: This Google Drive contains all the raw traffic captured from our testbeds that have been used in the paper. \href{https://drive.google.com/drive/folders/1Okww8Kcfwc7UB1ThD4XdB4gzWrWVLBSX?usp=sharing}{Link to dataset}
% \end{itemize}

\subsubsection{Hardware dependencies}
The S-SWaT implementations, ICS-SNIPER-attack and SOTA-detectors-NDSS require no specialized hardware or GPU. For \textit{S-SWaT}, we used three Ubuntu 22.04  instances on Amazon EC2: one simulating the compromised router (1 vCPU, 1 GB RAM, 32 GB storage), one simulating the cloud-hosted SCADA (4 vCPUs, 16 GB RAM, 32 GB storage), and one simulating the plant site that runs the six simulated PLCs (8 vCPUs, 32 GB RAM, 32 GB storage). All VMs must use Python 3.10 (for Modbus) or Python 2.7 (for ENIP), with OpenVPN and Easy-RSA installed on them. The VM simulating the compromised router should have tshark installed on it. We evaluated \textit{ICS-SNIPER-attack} on a Dell XPS 15 laptop equipped with a 12th Gen Intel Core i7 processor and 32 GB RAM. Other general-purpose computers with similar memory and storage configurations should suffice. For \textit{SOTA-detectors-NDSS}, we used a 2021 Apple MacBook Pro equipped with an Apple M1 Pro processor and 16 GB RAM. Other general-purpose computers with similar memory and storage configurations should suffice. 

\emph{H-SWaT (\texttt{Modbus-C-hardware-testbed})} requires six Programmable Logic Controllers (PLCs) instead of an Amazon EC2 instance. We used one Allen-Bradley Micro820 2080-LC20-20AWB PLC (rev.\ 14.11) and five Micro820 2080-LC20-20QWB PLCs (rev.\ 14.4); one Dell XPS 13 Plus 9320 laptop (Windows 11) acting as the plant gateway, process simulator, and Connected Components Workbench engineering workstation; one 1~Gbps USB Ethernet adapter linking the gateway to the PLC network; one Ubuntu 22.04 EC2 VM (1 vCPU, 1~GB RAM, 32~GB storage) simulating the compromised router; and one Ubuntu 22.04 VM (4 vCPUs, 16~GB RAM, 32~GB storage) to simulate the SCADA. Reviewers without this hardware can still validate H-SWaT's attack and detection results from the included traces.

\begin{table}[t]
\renewcommand{\tabcolsep}{4pt}
\begin{tabularx}{\columnwidth}{l>{\raggedright\arraybackslash}X}
\toprule
\rowcolor[HTML]{EFEFEF}
\textbf{Artifact} &
\textbf{Description}\\
\midrule
\href{https://github.com/ICS-Sniper/Modbus-C-hardware-testbed.git}{\texttt{Modbus-C-hardware-testbed}} &
H-SWaT testbed using the Connected Modbus protocol\\\midrule
\href{https://github.com/ICS-Sniper/Modbus-U-software-testbed.git}{\texttt{Modbus-U-software-testbed}} &
S-SWaT testbed using the Unconnected Modbus protocol\\\midrule
\href{https://github.com/ICS-Sniper/ENIP-U-software-testbed.git}{\texttt{ENIP-U-software-testbed}} &
S-SWaT testbed using the Unconnected ENIP protocol\\\midrule
\href{https://github.com/ICS-Sniper/ICS-SNIPER-attack.git}{\texttt{ICS-SNIPER-attack}} &
Implementation of \sniper{}\\\midrule
\href{https://github.com/ICS-Sniper/SOTA-detectors-NDSS.git}{\texttt{SOTA-detectors-NDSS}} &
IPAL detectors adapted to ICS traffic, preprocessed datasets, and evaluation and plotting code\\\midrule
\href{https://drive.google.com/drive/folders/1Okww8Kcfwc7UB1ThD4XdB4gzWrWVLBSX?usp=sharing}{\texttt{Dataset}} &
Raw traffic captured from our testbeds and used in the paper (Google Drive)\\
\bottomrule
\end{tabularx}
\caption{List of ICS-Sniper artifacts. Each artifact name links to its source. All repositories are hosted at \url{https://github.com/ICS-Sniper}.}
\label{tab:artifacts}
\end{table}
\renewcommand{\tabcolsep}{6pt}

\subsubsection{Software dependencies}
S-SWaT uses Python 3.10 on Ubuntu 22.04, with OpenVPN and Easy-RSA on the router and the OpenVPN client on the SCADA and PLC hosts. ICS-Sniper uses \texttt{tshark} for packet capture and Python for the profiling and packet-dropping modules. H-SWaT additionally requires Rockwell Automation Connected Components Workbench (to compile and deploy the structured-text control logic on the Micro820 PLCs) and the Windows plant simulator; the OpenVPN/Ubuntu stack for its router and SCADA matches S-SWaT. The defense artifact uses Python 3.12.2 (setup scripts support Python 3.9--3.12).

\subsubsection{Benchmarks}
The evaluation uses the six-stage SWaT plant control logic and physical-process models (adapted from the open-source single-host SWaT simulation) and the encrypted traffic traces captured from S-SWaT and H-SWaT for the seven configurations of Table~II (S-MD-BASE, S-EN-BASE, S-EN-RNC, S-EN-PCP, H-MD-BASE, H-MD-LOSS1, H-MD-LOSS2) and a 12-PLC-version of S-EN-PCP. It also contains encrypted traffic traces and SCADA log files captured during the attack. The defense artifact uses preprocessed IPAL datasets derived from the SCADA process logs and traffic traces of these runs. All required data are included; no external datasets are required.

\subsection{Artifact Installation \& Configuration}
Full step-by-step instructions for installation and configuration of each artifact are in each repository's README.

%%%%%%%%%%%%%%%%%%%%%%%%%%%%%%%%%%%%%%%%%%%%%%%%%%%%%%%%%%%%%%%%%%%%%%%%%%%%%%%%%%%%%%%%%%%%%%%%

\subsection{Experiment Workflow}

The end-to-end workflow has four steps.
\begin{enumerate}
    \item \emph{Testbed setup and traffic capture:} set up a testbed and run the process for at least one complete operational cycle while ICS-Sniper captures the encrypted VPN traffic at the compromised router.
    \item \emph{Process profiling:} run the ICS-Sniper process-profiling module on the capture to estimate the superperiod $L$ and rank the critical superperiods. 
    \item \emph{Active attack:} replay another cycle on the testbed, drop payload packets in the two targeted superperiods, log the resulting process behavior at the SCADA, capture encrypted traffic at the compromised router and unencrypted traffic at the SCADA, and measure the attack impact. 
    \item \emph{Evaluate IPAL detectors:} run the IPAL ensemble over the network and process logs to measure detection lag and alert overlap. 
\end{enumerate}
The included network traces and log files would allow reviewers to run Steps 2 -- 4 without running the testbed.

\subsection{Major Claims}
The artifacts support the paper's following four key results about ICS-Sniper.
\begin{enumerate}[label=\textbf{Claim-\arabic*}:, wide, labelwidth=!, labelindent=13pt]
    \item ICS-Sniper identifies the superperiod with 100\% accuracy across 8 testbed configurations discussed in \S~VI of the paper. This includes seven testbed configurations in Table~II as well as the testbed configuration with 12 PLCs (see \S~VI-A, supplementary evaluation (ii)), from encrypted-traffic metadata alone. The experiments have been described in \S~VI-A, and the results have been reported in Table~V.
    \item ICS-Sniper correctly identifies the critical superperiods, ranking the true first critical superperiod first in 71\% of cases and second in the remaining ones. The experiments have been described in \S VI-A, and the results have been reported in Table~VI.
    \item Dropping payload packets in the two targeted superperiods disrupts critical state transitions and degrades plant throughput. The Process Delay Attack postpones clean-water pumping by $\approx$15~min (S-EN-BASE) and 2~min (H-MD-BASE), and the UF-Depletion Attack leaves the P3 backwash valve open for 9~min~11~s after the P6 pump stops, while dropping only $\approx$2.5--4.7\% of total packets. The experiments have been described, and the results have been reported in \S VI-B.
    \item No detector in IPAL effectively detects ICS-Sniper's active attack: the ALL ensemble never fires during the attack, in 2 of 5 scenarios no detector fires during the attack, and process degradation always precedes any alert. The experiments have been described, and the results have been reported in \S VI-C, and Figs.~9--12 of Appendix~\ref{subsec:alerts}.
\end{enumerate}

\subsection{Evaluation}
 
\subsubsection{Experiment (E1) to verify Claim-1: Superperiod identification}
[30 human-minutes + $\approx$8 compute-hours (can be cut down to $\approx$1 compute-hour by avoiding the metadata extraction step; metadata file is already provided in the ICS-SNIPER-attack repository)]: run the profiling module on each configuration and compare the estimated superperiod to ground truth.
 
\textit{[Preparation]} Select an included trace (or run a testbed cycle) for each of the seven configurations of Table~II.
 
\textit{[Execution]} Run the ICS-Sniper profiling module ($\phi_{\text{overlap}}$, $\phi_{\text{interval}}$, $\phi_{\text{pktlen}}$) on each trace.
 
\textit{[Results]} The estimated superperiod should match the actual value for every configuration (accuracy $=100\%$), reproducing Table~V. 
 
\subsubsection{Experiment (E2) to verify Claim-2: Critical superperiod ranking}
[15 human-minutes + $<$1 compute-hour]: run the Step-4 heuristics to rank critical superperiods.
 
\textit{[Preparation]} Use the superperiod values from (E1).
 
\textit{[Execution]} Run the critical-superperiod ranking on each configuration.
 
\textit{[Results]} The identified critical transitions should match Table~VI.
 
\subsubsection{Experiment (E3) to verify Claim-3: Attack impact}
[30 human-minutes + $\approx$4-5 compute-hours per full cycle (Can be cut down to $<1$ compute-hour by using the provided network traces and logs)]: execute the Process Delay and UF-Depletion attacks and measure process impact.
 
\textit{[Preparation]} Configure S-EN-BASE and H-MD-BASE; enable the active-attack (packet-dropping) module targeting the two superperiods selected in (E2).
 
\textit{[Execution]} Run one operational cycle per attack while the SCADA logs sensor/actuator values.
 
\textit{[Results]} The Process Delay Attack should postpone clean-water pumping by $\approx$15~min (S-EN-BASE) and 2~min (H-MD-BASE, plus $\approx$3~min extra operation); the UF-Depletion Attack should hold the P3 backwash valve open $\approx$9~min~11~s after the P6 pump stops; dropped packets total $\approx$2.5--4.7\% of the cycle. Minor timing variation is expected due to network jitter; the qualitative disruption and order-of-magnitude figures should hold.
 
\subsubsection{Experiment (E4) to verify Claim-4: Detector evasion}
[30 human-minutes + $\approx$1--2 compute-hours]: run the IPAL ensemble over attack logs and measure detection lag and alert overlap.
 
\textit{[Preparation]} Collect the network and process logs from (E3) (or use the included logs) for the five scenarios of \S VI-C.
 
\textit{[Execution]} Run \texttt{SOTA-detectors-NDSS} (individual detectors and the ANY/ALL ensembles) and the plotting utilities.
 
\textit{[Results]} Reproduce Table~III and Figs.~9--12: no network-based detector fires before the attack concludes, the ALL ensemble never fires during the attack, and in 2 of 5 scenarios no detector fires during the active attack stage.